\documentclass[12pt]{article}

\usepackage[margin=1in]{geometry}

\usepackage{enumitem} 

\usepackage{centernot}

\usepackage{times}

\usepackage{setspace} 
\usepackage{caption}
\usepackage{amsmath}
\usepackage{amssymb}
\usepackage{amsthm} % \theoremstyle{style}
\usepackage{mathtools} % matrix*: [c] centered, [l] left-aligned, [r] right-aligned

\usepackage[utf8]{inputenc} 
\usepackage[english]{babel}

\usepackage{graphicx} 

\usepackage{float}

\usepackage{siunitx} 
\newtheorem{assumption}{Assumption}
\newtheorem{proposition}{Proposition}
\newtheorem{lemma}{Lemma}

\usepackage[title]{appendix}

\numberwithin{equation}{section}

\newcommand\blfootnote[1]{
  \begingroup
  \renewcommand\thefootnote{}\footnote{#1}
  \addtocounter{footnote}{-1}
  \endgroup
}

\usepackage[hang]{footmisc}
\usepackage{hyperref}
\hypersetup{colorlinks=true, allcolors=black}

\usepackage[authoryear,round,sort,comma,longnamesfirst]{natbib} 
\usepackage{myapalike}
\begin{document}

\begin{titlepage}
\begin{center}
\linespread{1.2}
\vspace*{1cm} % Remove if acknowledgements are included
\Large{\textbf{Structural Gaussian mixture vector autoregressive model with application to the asymmetric effects of monetary policy shocks}}\\

\vspace{1cm} 
\Large{Savi Virolainen}\\
\large{University of Helsinki}\\
\vspace{1cm}

\begin{abstract}
\noindent %A structural Gaussian mixture vector autoregressive model is introduced. The shocks are identified by combining simultaneous diagonalization of the reduced form error covariance matrices with constraints on the time-varying impact matrix. This leads to more flexible identification conditions than in the conventional SVAR models, and some of the constraints are also testable. In an empirical application to quarterly U.S. data covering the period from 1953Q3 to 2021Q4, our model identifies two regimes: a stable inflation regime and an unstable inflation regime. The unstable inflation regime is characterized by high or volatile inflation, and it mainly prevails in the 1970's, early 1980's, during the Financial crisis, and in the COVID-19 crisis from 2020Q3 onwards. The stable inflation regime, in turn, is characterized by moderate inflation, and it prevails when the unstable inflation regime does not. While the effects of the monetary policy shock are relatively symmetric in the unstable inflation regime, we find strong asymmetries with respect to the sign and size of the shock as well as to the initial state of the economy in the stable inflation regime. On average, the real effects of the monetary policy shock are somewhat stronger in the stable inflation regime than in the unstable inflation regime. The accompanying, CRAN distributed R package gmvarkit provides a comprehensive set of tools for numerical analysis of the model.\\[1cm]
% 100 word abstract:
A structural Gaussian mixture vector autoregressive model is introduced. The shocks are identified by combining simultaneous diagonalization of the reduced form error covariance matrices with constraints on the time-varying impact matrix. This leads to flexible identification conditions, and some of the constraints are also testable. The empirical application studies asymmetries in the effects of the U.S. monetary policy shock and finds strong asymmetries with respect to the sign and size of the shock and to the initial state of the economy. The accompanying CRAN distributed R package gmvarkit provides a comprehensive set of tools for numerical analysis.\\[1cm]

\center{\textbf{Acknowledgements}}\\
This work was supported by the Academy of Finland under Grant 308628. The author thanks Markku Lanne, Mika Meitz,  and Pentti Saikkonen who helped to improve this paper substantially. The author also thanks Henri Nyberg and Antti Ripatti for the useful comments.\\[1cm]

%[2.0cm]
\end{abstract}

\vfill

%\blfootnote{This work was supported by the Academy of Finland under Grant 308628. The author thanks Markku Lanne, Mika Meitz,  and Pentti Saikkonen who helped to improve this paper substantially. The author also thanks Henri Nyberg and Antti Ripatti for the useful comments.}
\blfootnote{Contact address: Savi Virolainen, Faculty of Social Sciences, University of Helsinki, P. O. Box 17, FI–00014 University of Helsinki, Finland; e-mail: savi.virolainen@helsinki.fi.}
\blfootnote{The author has no conflict of interest to declare.}

\end{center}
\end{titlepage}

\section{Introduction}
Tracing out the effects of an economic shock is a major task in econometrics. A popular approach is to consider a set of key variables and utilize a structural vector autoregressive (SVAR) or structural error correction (SVEC) model for the purpose. They have well established theoretical grounds \citep[see][and the references therein]{Kilian+Lutkepohl:2017} and are accommodated by many of the popular statistical software packages. Linear SVAR and SVEC models are not, however, suitable for modelling series in which the underlying data generating dynamics are nonlinear or the shocks have asymmetric effects in different states of the economy. Models capable of capturing such features include mixture models, such as the mixture vector autoregressive model \citep{Fong+Li+Yau+Wong:2007}, the mixture periodic vector autoregressive model \citep{Bentarzi+Djeddou:2014}, the Gaussian mixture vector autoregressive (GMVAR) model \citep{Kalliovirta+Meitz+Saikkonen:2016}, and the logit mixture vector autoregressive model \citep{Burgard+Neuenkirch+Nockel:2019}.

This paper introduces a structural version of the GMVAR model. In the structural GMVAR (SGMVAR) model of autoregressive order $p$, the regime-switching dynamics are endogenously determined by the full distribution of the previous $p$ observations. Specifically, the greater the relative weighted likelihood of a regime is, the more likely the process is to generate an observation from it. This facilitates associating statistical characteristics and economic interpretations to the regimes. The SGMVAR model thus allows the regime-switches to depend on a richer set of statistical characteristics of the data than many of the popular threshold VAR \citep{Tsay:1998} and smooth transition VAR \citep{Anderson+Vahid:1998} models in which regime-transitions often depend only on the level of the transition-variables. The specific formulation of the mixing weights also leads to attractive theoretical properties, such as ergodicity and fully known stationary distribution of $p+1$ consecutive observations. %, which is exceptional for a nonlinear VAR model (täsmennä että fully known stationary distribution). %The knowledge of the stationary distribution is exceptional for a nonlinear VAR model, and can be taken advantage of in the estimation of the (generalized) impulse response function.

The effects of the structural shocks depend on the initial values of the included variables, and they are also allowed to vary according to the sign and size of the shock due to possibly resulting regime-switches. Consequently, the (generalized) impulse response functions reflect the prevailing macroeconomic conditions that are transmitted to the regime-switching probabilities through the level, variability, and temporal as well as contemporaneous dependence of the past observations. Because the shocks may have asymmetric effects with respect to their size, the conditional heteroskedasticity of the reduced form error needs to be controlled for. Therefore, the impact matrix of the SGMVAR model is time-varying and constructed so that it captures the conditional heteroskedasticity of the reduced form error, thereby enabling to standardize the conditional variance of each structural shock to a constant. The initial effects of a constant-sized structural shock are, hence, amplified according to the conditional variance of the reduced form error, also reflecting the prevailing state of the economy.

Identification of the shocks requires that they are simultaneously orthogonalized in all regimes. We show that  together with any constant standardization of the structural shock's conditional variance, this condition generally leads to a unique identification of the impact matrix up to ordering of its columns and changing all signs in a column. Thus, as long as one is willing to impose the assumption of a single (time-varying) impact matrix, the columns of the impact matrix unambiguously characterize the estimated impact effects of the shocks without further constraints. The identification does not, however, reveal which column of the impact matrix is related to which shock. Since the impact matrix is also subject to estimation error, further constraints may be needed for labelling the shocks. The constraints are testable, as they are overidentifying.

In order to formulate the impact matrix and the identification conditions, it is convenient to utilize the well known matrix decomposition \citep[Theorem A9.9]{Muirhead:1982} proposed by \cite{Lanne+Lutkepohl:2010} and \cite{Lanne+Lutkepohl+Maciejowska:2010} for a similar identification problem. \cite{Lanne+Lutkepohl:2010} assume that the reduced form error covariance matrices admit this decomposition, then show that the shocks are statistically identified, and finally test conventional zero constraints that lead to economically interpretable shocks. \cite{Lanne+Lutkepohl+Maciejowska:2010}, in turn, note that the shocks are readily identified when the matrix decomposition is imposed to the reduced form error covariance matrices. Our approach differs from them in that we obtain locally identified structural shocks by directly investigating the properties of the impact matrix. We also provide a general set of conditions for identifying any subset of the shocks that allows for using sign constraints alone or together with zero constraints. Moreover, we (partially) relax a technical condition required for statistical identification of the model and allow identification of a subset of the shocks when the model is only partially identified. 
 
Our empirical application studies asymmetries in the expected effects of monetary policy shocks in the U.S. using a quarterly series covering the period from 1954Q3 to 2021Q4. Our SGMVAR model identifies two regimes: a stable inflation regime and an unstable inflation regime. The unstable inflation regime is characterized by high or volatile inflation, and it mainly prevails in the 1970's, early 1980's, during the Financial crisis, and in the COVID-19 crisis from 2020Q3 onwards. The stable inflation regime, in turn, is characterized by moderate inflation, and it prevails when the unstable inflation regime does not. We find the effects of the monetary policy shock relatively symmetric in the unstable inflation regime, as it rarely causes a switch to the stable inflation regime. A contractionary (expansionary) monetary policy shock appears to first increase (decrease) inflation after which the inflation significantly decreases (increases) for several years. The strong contraction (expansion) in the cyclical component of the GDP lasts for roughly three years and is followed by a small short-term expansion (contraction) before the response decays to zero. 

In the stable inflation regime, the (generalized) impulse responses are strongly asymmetric with the respect to the sign and size of the monetary policy shock as well as to the initial state of the economy. A contractionary shock causes, on average, roughly a three-year hump-shaped contraction of the GDP, but it also seems to increase inflation by driving the economy towards the unstable inflation regime. A small expansionary shock does not move prices much on average, but a large expansionary shock often drives the economy towards the unstable inflation regime and propagates high and persistent inflation. The high inflation is followed by a significant monetary policy tightening and a persistent contraction of the GDP after the initial expansion. On average, the real effects of the monetary policy shock are found somewhat stronger in the stable inflation regime than in the unstable inflation regime.
 
The GMVAR model has been previously applied in impulse response analysis by \cite{Kalliovirta+Malinen:2020}, who %(study cross-country dependence of income inequality and) 
identify the shocks by constraining the reduced form error covariance matrices and allow the impact responses of the variables to vary relative to each other across the regimes. Our assumption of a common (time-varying) impact matrix for all the regimes constraints the relative magnitudes of the impact responses of the variables to be time-invariant (for each shock), but it leads to flexible identification conditions and enables to test the validity of the identifying constraints. %Unlike \cite{Kalliovirta+Malinen:2020}, we also allow the regime to switch as a result of a shock and estimate the true (generalized) impulse response functions of a non-linear VAR. 
%implying that the varying economic conditions are appropriately taken into account in the (generalized) impulse response functions through the regime-switching mechanics.
\cite{Kalliovirta+Malinen:2020} estimate the impulse response functions for each regime of the GMVAR model separately as if each of them was a linear VAR. In contrast, we allow the regime to switch as a result of a shock and estimate the true (generalized) impulse response functions of the non-linear VAR.
%Moreover, the applied GMVAR model in \cite{Kalliovirta+Malinen:2020} constraints the autoregression matrices to be equal across the regimes, while we do not impose such constraints (equality of the AR matrices and intercepts across the regimes is, however, tested in Appendix~\ref{sec:details}).
%"Jos impulssivasteet todella ovat regiimikohtaisia eli olettavat regiimin pysyvän samana, kannattaa tuoda esiin, että sinun paperissasi ne ovat aitoja epälineaarisen mallin mukaisia impulssivasteita"

Structural mixture VARs, in general, have been previously applied for studying to the effects of monetary policy shocks at least by \cite{Burgard+Neuenkirch+Nockel:2019}, who proposed a mixture VAR with logistic mixing weights and Cholesky identified shocks. As opposed to \cite{Burgard+Neuenkirch+Nockel:2019}, our identification scheme is more flexible in the sense that it does not require many (or necessarily any) zero constraints on the impact effects of the shocks. Moreover, in our model the regime-switching probabilities depend on the full distribution of the preceding $p$ observations instead of just on the level of the switching-variables. 
 
The rest of this paper is organized as follows. Section~\ref{sec:gmvar} defines the reduced form GMVAR model. In Section~\ref{sec:sgmvar}, the structural GMVAR model is first introduced. Then, identification of the shocks and estimation of the model parameters are discussed. Section~\ref{sec:girf} discusses impulse response analysis and describes the generalized impulse response function (GIRF) \citep{Koop+Pesaran+Potter:1996}. Section~\ref{sec:empapp} presents the empirical application and Section~\ref{sec:summary} summarizes. Appendices provide proofs for the stated lemma and propositions, a Monte Carlo algorithm for estimating the GIRF, and details on the empirical application. Finally, we have accompanied this paper with the CRAN distributed R package gmvarkit \citep{gmvarkitnormal}, which is comprehensively documented and provides a comprehensive set of tools for numerical analysis of the model.

\section{Reduced form GMVAR model}\label{sec:gmvar}
To build theory and notation, consider first the reduced form GMVAR model introduced by \cite{Kalliovirta+Meitz+Saikkonen:2016}. Let $y_t$ ($t=1,2,...$) be the $d$-dimensional time series of interest and $\mathcal{F}_{t-1}$ denote the $\sigma$-algebra generated by the random vectors $\lbrace y_{t-j}, j>0 \rbrace$. For a GMVAR model with $M$ mixture components and autoregressive order $p$, we have 
\begin{align}
y_t &=\sum_{m=1}^M s_{m,t}(\mu_{m,t} + u_{m,t}), \quad u_{m,t} \sim NID(0, \Omega_m) \label{eq:gmvar1}\\
\mu_{m,t} &= \phi_{m,0} + \sum_{i=1}^{p}A_{m,i}y_{t-i}, \quad m=1,...,M,\label{eq:gmvar2}
\end{align}
where $\phi_{m,0}\in\mathbb{R}^{d}$ are intercept parameters,  $\Omega_m$ are positive definite covariance matrices, and for each $m$, the coefficient matrices $A_{m,i}$, $i=1,...,p$, are assumed to satisfy the usual stability condition
\begin{equation}
\det\left(I_d - \sum_{i=1}^p A_{m,i}z^i\right)\neq 0 \ \ \text{for} \ \ |z|\leq 1, \ \ m=1,...,M,
\end{equation}
which guarantees stationarity of the component processes. The unobservable regime variables $s_{1,t},...,s_{M,t}$ are such that at each $t$, exactly one of them takes the value one and the others take the value zero according to the conditional probabilities $\mathrm{P}(s_{m,1}=1|\mathcal{F}_{t-1})\equiv\alpha_{m,t}$ that satisfy $\sum_{m=1}^{M}\alpha_{m,t}=1$. The normally and independently distributed (NID) errors $u_{m,t}$ are assumed independent of $\mathcal{F}_{t-1}$, and conditional on $\mathcal{F}_{t-1}$, $(s_{1,t},...,s_{M,t})$ and $u_{m,t}$ are independent. 

The definition (\ref{eq:gmvar1})-(\ref{eq:gmvar2}) implies that at each $t$, the process generates an observation from one of its mixture components, a linear VAR process, that is randomly selected according to the probabilities given by the mixing weights $\alpha_{m,t}$. Denoting $\boldsymbol{y}_{t-1}=(y_{t-1},...,y_{t-p})$, the mixing weights are defined as \citep[Equation~(7)]{Kalliovirta+Meitz+Saikkonen:2016}
\begin{equation}\label{eq:alpha_mt}
\alpha_{m,t} = \frac{\alpha_m n_{dp}(\boldsymbol{y}_{t-1};\boldsymbol{1}_p\otimes \mu_m, \boldsymbol{\Sigma}_m)}{\sum_{n=1}^M \alpha_n n_{dp}(\boldsymbol{y}_{t-1};\boldsymbol{1}_p\otimes \mu_n, \boldsymbol{\Sigma}_n)}, \ \ m=1,...,M,
\end{equation}
where $\alpha_1,...,\alpha_M$ are mixing weight parameters that satisfy $\sum_{m=1}^M \alpha_m=1$ and $n_{dp}(\cdot;\boldsymbol{1}_p\otimes \mu_m, \boldsymbol{\Sigma}_m)$ is the density function of the $dp$-dimensional normal distribution with mean $\boldsymbol{1}_p\otimes \mu_m$ and covariance matrix $\boldsymbol{\Sigma}_m$. The symbol $\boldsymbol{1}_p$ denotes a $p$-dimensional vector of ones, $\otimes$ is Kronecker product, $\mu_m=(I_d - \sum_{i=1}^pA_{m,i})^{-1}\phi_{m,0}$, and the covariance matrix $\boldsymbol{\Sigma}_m$ is given in \cite{Lutkepohl:2005}, Equation~(2.1.39), but using the parameters of the $m$th component process. That is, $n_{dp}(\cdot;\boldsymbol{1}_p\otimes \mu_m, \boldsymbol{\Sigma}_m)$ corresponds to the density function of the stationary distribution of the $m$th component process.

The mixing weights are thus weighted ratios of the component process stationary densities corresponding to the preceding $p$ observations. This implies that the greater the weighted relative likelihood of a regime is, the more likely the process is to generate an observation from it. This facilitates associating statistical characteristics and economic interpretations to the regimes. In addition to the (generalized) impulse response functions of the observable variables, the responses of the mixing weights may therefore be of interest. The definition of the mixing weights also leads to attractive theoretical properties such as ergodicity and full knowledge of the stationary distribution of $p+1$ consecutive observations \citep[Theorem~1, see the proof of Theorem~1 for the stationary distribution of $p+1$ consecutive observations]{Kalliovirta+Meitz+Saikkonen:2016}. Specifically, the stationary distribution of the process $\boldsymbol{y}_{t}=(y_{t},...,y_{t-p+1})$ is a mixture of $dp$-dimensional normal distributions that is characterized by the density
\begin{equation}
f(\boldsymbol{y})=\sum_{m=1}^M \alpha_m n_{dp}(\boldsymbol{y};\boldsymbol{1}_p\otimes \mu_m, \boldsymbol{\Sigma}_m).
\end{equation}
The knowledge of the stationary distribution is taken advantage of in the impulse response analysis in Section~\ref{sec:girf} and Appendix~\ref{sec:montecarlo}.

\section{Structural GMVAR model}\label{sec:sgmvar}

\subsection{The model setup}\label{sec:tworegime}
Consider the GMVAR model defined in (\ref{eq:gmvar1})-(\ref{eq:gmvar2}).  We focus on the "B-model" setup and write the structural GMVAR model as 
\begin{equation}\label{eq:sgmvar}
y_t = \sum_{m=1}^M s_{m,t}\left(\phi_{m,0} + \sum_{i=1}^{p}A_{m,i}y_{t-i}\right) + B_te_t,
\end{equation}
and
\begin{equation}\label{eq:sgmvarerr}
u_t \equiv B_te_t = 
\left\lbrace\begin{matrix} 
u_{1,t}\sim N(0,\Omega_1) & \text{if} & s_{1,t}=1 & (\text{with probability } \alpha_{1,t}) \\
u_{2,t}\sim N(0,\Omega_2) & \text{if} & s_{2,t}=1 & (\text{with probability } \alpha_{2,t}) \\
\vdots & & & \\
u_{M,t}\sim N(0,\Omega_M) & \text{if} & s_{M,t}=1 & (\text{with probability } \alpha_{M,t}) 
\end{matrix}\right. 
\end{equation}
where the probabilities are expressed conditionally on $\mathcal{F}_{t-1}$ and $e_t$ is an orthogonal structural error. Unlike in the conventional SVAR analysis, the invertible $(d\times d)$ "B-matrix" (or impact matrix) $B_t$, which governs the contemporaneous relations of the shocks, is time-varying and a function of $y_{t-1},...,y_{t-p}$. This enables to amplify a constant-sized structural shock according to the conditional variance of the reduced form error, which varies according to the mixing weights.  Appropriate modelling of conditional heteroskedasticity in the B-matrix is of interest, because the (generalized) impulse response functions may be asymmetric with respect to the size of the shock. 

We have $\Omega_{u,t}\equiv\text{Cov}(u_t|\mathcal{F}_{t-1})=\sum_{m=1}^M\alpha_{m,t}\Omega_m$, while the conditional covariance matrix of the structural errors $e_t=B_t^{-1}u_t$ (which have a mixture normal distribution and are not IID but martingale differences and therefore uncorrelated) is obtained as 
\begin{equation}\label{eq:condcovmat}
\text{Cov}(e_t|\mathcal{F}_{t-1})=\sum_{m=1}^M\alpha_{m,t}B_t^{-1}\Omega_mB_t'^{-1}.
\end{equation}
The B-matrix $B_t$ should therefore be chosen so that the structural shocks are orthogonal regardless of which regime they come from. We will next discuss the properties of any such B-matrix that solves the diagonalization problem. Then, we present a locally unique solution under a constant normalization of the structural error's conditional variance. After that, in the following two subsections, we will discuss global identification of the shocks, allowing also only partial identification of the model.%Viimeinen virke toiseen kohtaan?

% TÄHÄNKÖ VÄLIIN SELITYS SIITÄ MIKÄ ON TILASTOLLINEN IDENTIFIOINTI LÄHTEINEEN?
%...In other words/That is, we show that the shocks are statistically identified under a technical condition (Assumption~\ref{as:eigenvalues}} below) without further constraints on model (SEE EG LÄHDE FOR DISCUSSION ON STATISTICAL IDENTIFICATION). 

Specifically, we show that our model readily identifies the B-matrix up to ordering of its columns and changing all signs in a column, %through changes in the conditional volatility, %without further constraints on the model, 
but it is not revealed which column of the B-matrix is related to which shock. The identification follows from the assumption $e_t=B_t^{-1}u_t$, which (as we show in this section) implies that for each shock the relative magnitudes of the impact responses of the variables stay constant over time.\footnote{See \citet[Chapter~14]{Kilian+Lutkepohl:2017} for a discussion on identification by heteroskedasticity in a linear VAR model.}
% For an up-to-date discussion on identification through heteroskedasticity, we refer to \cite[Chapter~14 and the references therein]{Kilian+Lutkepohl:2017}.
%
%This is different to the conventional SVAR identification, where economically motivated constraints are imposed on model foer
%
This is different to the conventional SVAR setup, where the identification of the B-matrix requires further constraints to be imposed on model. Conventionally, the shocks are often identified, for instance, by placing economically motivated zero constraints on the impact or the long-run effects of the shocks \citep[e.g.,][Chapters~8 and 10]{Kilian+Lutkepohl:2017}. Sign constraints, in turn, are commonly used to obtain a set identification with less restrictive or economically more plausible constraints \citep[e.g.,][Chapter~13]{Kilian+Lutkepohl:2017}.
%
%After discussing the local uniqueness of the B-matrix, we will discuss global identification of the shocks in the following two subsections, allowing also only partial identification of the model.%Viimeinen virke toiseen kohtaan?
%In order to formally label the
%

In Section~\ref{sec:identshocks}, also we make use of zero and sign constraints, but we do it in order to formally label the already locally identified columns of the B-matrix by the shocks of interest. 
%but we do it in order to formally label the correct column(s) of the B-matrix by the shock(s) of interest and to obtain globally identified shocks based on the local solution (that is presented later in this section).
The required conditions are, nevertheless, flexible, and allow for using sign constraints alone or together with zero constraints. Some of the constraints are also testable, as they are overidentifying. %which is a major advantage of our approach compared to the conventional identification methods. 
Section~\ref{sec:partial_ident} additionally takes advantage of zero constraints to identify the shock of interest when the condition for identification through conditional heteroskedasticity fails.%TÄMÄN TYYPPISEN TEKSTIN VOISI TOKI LAITTAA KENTIES VAIN MYÖS IDENTIFIOINTISECTIONIIN?

It turns out that any invertible B-matrix that simultaneously diagonalizes the covariance matrices $\Omega_1,...,\Omega_M$, i.e., produces a diagonal conditional covariance matrix (\ref{eq:condcovmat}) of the structural error, has linearly independent eigenvectors of the matrix $\Omega_m\Omega_1^{-1}$ as its columns. If $M > 2$, the matrices $\Omega_m\Omega_1^{-1}$, $m=2,...,M$, thus need to share the common eigenvectors in $B_t$, which restricts the parameter space for the covariance matrices. In that case, the existence of such B-matrix can be tested with a likelihood ratio test, for example. Denoting the eigenvalues of $\Omega_m\Omega_1^{-1}$ as $\lambda_{mi}$, the B-matrix is also unique up to scalar multiples and ordering of its columns if none of the pairs of $\lambda_{mi}$, $i=1,...,d$, is identical for all $m=2,...,M$. These results are formalized in the following assumption and lemma. 
 
\begin{assumption}\label{as:eigenvalues}
Consider $M$ positive definite $(d\times d)$ covariance matrices $\Omega_m$,  $m = 1,...,M$, and denote the strictly positive eigenvalues of the matrices $\Omega_m\Omega_1^{-1}$ as $\lambda_{mi}$, $i=1,...,d$,  $m=2,...,M$.  Suppose that for all $i\neq j\in\lbrace 1,...,d\rbrace$, there exists an $m\in\lbrace 2,...,M\rbrace$ such that $\lambda_{mi}\neq\lambda_{mj}$.
\end{assumption}

\begin{lemma}\label{lemma1}
Consider $M$ positive definite $(d\times d)$ covariance matrices $\Omega_m$,  $m = 1,...,M$, and an invertible $(d\times d)$ matrix $B_t$ such that $B_t^{-1}\Omega_m B_t'^{-1}$ are diagonal matrices with strictly positive diagonal elements. Then, $B_t$ has eigenvectors of $\Omega_m\Omega_1^{-1}$ as its columns. Moreover, $B_t$ is unique up to scalar multiples and ordering of its columns if Assumption~\ref{as:eigenvalues} holds.
\end{lemma}

Under Assumption~\ref{as:eigenvalues}, the columns of $B_t$ are unique up to scalar multiples and ordering, implying that the shocks are identified up to sign, size, and ordering. Normalizing the conditional covariance matrix of the structural error to a constant diagonal matrix then identifies the B-matrix up to sign and ordering of the shocks. This is formalized in the following proposition.

\begin{proposition}\label{prop:Bmat_uniqueness}
Consider $M$ positive definite $(d\times d)$ covariance matrices,  $\Omega_m$,  $m=1,...,M$, and an invertible $(d\times d)$ matrix $B_t$ such that $B_t^{-1}\Omega_m B_t'^{-1}$  are diagonal matrices with strictly positive diagonal elements. Suppose that Assumption~\ref{as:eigenvalues} holds. Then, if the conditional covariance matrix of the structural error,  $\text{Cov}(e_t|\mathcal{F}_{t-1})=\sum_{m=1}^M\alpha_{m,t}B_t^{-1}\Omega_mB_t'^{-1}$, is normalized to a constant diagonal matrix with strictly positive diagonal entries,  the B-matrix $B_t$ is unique up to ordering of its columns and changing all signs in a column. 
\end{proposition}

That is, by fixing an ordering and signs for the columns of the B-matrix, the solution to the diagonalization problem is unique for any given (constant) normalization of the structural error's conditional covariance matrix, say, an identity matrix. In order to find the related B-matrix, it is then convenient to utilize the following matrix decomposition for the reduced form error covariance matrices, which was also employed by \cite{Lanne+Lutkepohl:2010} and \cite{Lanne+Lutkepohl+Maciejowska:2010} to solve a similar identification problem. We decompose the reduced form error covariance matrices as
\begin{equation}\label{eq:decomp}
\Omega_1 = WW' \quad \text{and} \quad \Omega_m=W\Lambda_mW', \quad m=2,...,M,
\end{equation}
where the diagonal of $\Lambda_m=\text{diag}(\lambda_{m1},...,\lambda_{md})$, $\lambda_{mi}>0$ ($i=1,...,d$), contains the eigenvalues of the matrix $\Omega_m\Omega_1^{-1}$ and the columns of the nonsingular $W$ are the related eigenvectors (that are the same for all $m$ by construction). When $M=2$, the decomposition (\ref{eq:decomp}) always exists \citep[Theorem A9.9]{Muirhead:1982}, but for $M>2$ its existence requires that the matrices $\Omega_m\Omega_1^{-1}$ share the common eigenvectors in $W$.  This is, however, testable and relates to our earlier discussion on the existence of a B-matrix that simultaneously diagonalizes the reduced form error covariance matrices.
 
Any scalar multiples of linearly independent eigenvectors of $\Omega_m\Omega_1^{-1}$ comprise an appropriate B-matrix, but only specific scalar multiples comprise the locally unique B-matrix associated with a given normalization of structural error's conditional covariance matrix. Direct calculation shows that the B-matrix associated with the normalization $\text{Cov}(e_t|\mathcal{F}_{t-1})=I_d$ is obtained as
\begin{equation}\label{eq:bt}
B_t=W(\alpha_{1,t}I_d + \sum_{m=2}^M\alpha_{m,t}\Lambda_m)^{1/2},
\end{equation}
where $B_tB_t'=\Omega_{u,t}$. Since $B_t^{-1}\Omega_mB_t'^{-1}= \Lambda_m \left( \sum_{n=1}^M\alpha_{n,t}\Lambda_n  \right)^{-1}$ where $\Lambda_1\equiv I_d$,  the B-matrix~(\ref{eq:bt}) simultaneously diagonalizes $\Omega_1,...,\Omega_M$, and $\Omega_{u,t}$ for each $t$ so that the structural error's conditional covariance matrix is normalized to an identity matrix:
\begin{equation}
\text{Cov}(e_t|\mathcal{F}_{t-1}) =  \sum_{m=1}^M\alpha_{m,t}\Lambda_m  \left( \sum_{n=1}^M\alpha_{n,t}\Lambda_n  \right)^{-1} = I_d.
\end{equation}
Our specification of the B-matrix differs from \cite{Lanne+Lutkepohl+Maciejowska:2010} who assume that the instantaneous effects of the shocks are time-invariant (and specify $B_t=W$), but it extends the one in \cite{Lanne+Lutkepohl:2010} to accommodate time-varying mixing weights.  

The SGMVAR model assumes a single B-matrix that varies continuously in time according to the conditional covariance matrix of the reduced form error, which in turn varies according to the mixing weights. We established that under Assumption~\ref{as:eigenvalues} and a normalization of the structural error's conditional variance, the B-matrix is unique up to ordering of its columns and switching all signs in a column. Hence, as long as one is willing to impose the assumption of a single (time-varying) B-matrix,  the columns of the B-matrix unambiguously characterize the estimated impact effects of the shocks, but they do not reveal which column is related to which shock. Since the impact matrix is also subject to estimation error, further constraints may be needed for labelling the shocks.\footnote{As opposed to our single B-matrix, an alternative specification of the structural model would incorporate a separate B-matrix for each of the regimes. Our B-matrix allows the magnitude of the impact effects of a constant sized shock to vary according to the mixing weights, but unlike our model, the alternative specification would allow variation also in the impact effects relative to the other variables. That is, our model imposes structure already in the model Equations~(\ref{eq:sgmvar}) and (\ref{eq:sgmvarerr}).}

\subsection{Identification of the shocks}\label{sec:identshocks}
We derived a locally unique solution for the B-matrix~(\ref{eq:bt}) under Assumption~\ref{as:eigenvalues}. However, global identification requires fixing the signs and the ordering of its columns. The signs can be fixed by placing a single strict sign constraint in each of the columns of $W$, whereas the ordering of the columns can be fixed by fixing an ordering for the eigenvalues $\lambda_{mi}$ in the diagonals of $\Lambda_m$. This leads to statistical identification of the model with any arbitrary ordering, but it does not reveal which column of the B-matrix is related to which shock. 

A structural shock relates to an economic shock through the specific constraints in the corresponding column of $W$ (or equally of the B-matrix) that only the shock of interest satisfies. If such constraints are readily satisfied in the (unrestricted) estimate of $W$, the identification amounts to labelling the structural shocks by the appropriate economic shocks, as long as the constraints are strong enough to pin down a unique ordering for the columns of $W$ (this argument will be formalized in Proposition~\ref{prop:ident1} below). If the unrestricted estimate of $W$ is such that the shocks of interest cannot be uniquely associated to it, the appropriate constraints can be placed for their identification.\footnote{For a more thorough discussion on economic shocks and their identification, see \cite{Ramey:2016} and \cite{Uhlig:2017}, for example.}

%The major difference to the conventional SVAR setup is that under Assumption~\ref{as:eigenvalues}, the B-matrix is already locally identified.

As in practice the interest is often in identifying only some specific shock or shocks, it is of interest to consider only partial identification of the B-matrix as well. Specifically, we say that the $j$th structural shock is uniquely identified if the $j$th column of the B-matrix~(\ref{eq:bt}) is unique for given mixing weights $\alpha_{1,t},...,\alpha_{M,t}$. This requires that the $j$th columns of $W$ and $\Lambda_m$, $m=2,..,M$, are unique. 
The following proposition gives sufficient conditions for global identification of the last $d_1$ shocks when the related pairs of $\lambda_{mi}$ are distinct for some $m$ (which is always the case under Assumption~\ref{as:eigenvalues} but does not require Assumption~\ref{as:eigenvalues} if $d_1<d$).

\begin{proposition}\label{prop:ident1}
Suppose $\Omega_1=WW'$ and $\Omega_m=W\Lambda_mW', \ \ m=2,...,M,$ where $\Lambda_m=\text{diag}(\lambda_{m1},...,\lambda_{md})$, $\lambda_{mi}>0$ ($i=1,...,d$), contains the eigenvalues of $\Omega_m\Omega_1^{-1}$ in the diagonal and the columns of the nonsingular $W$ are the related eigenvectors. Then, the last $d_1$ structural shocks are uniquely identified if
\begin{enumerate}[label={(\arabic*)}]
\item for all $j > d - d_1$ and $i\neq j$ there exists an $m\in \lbrace 2,...,M \rbrace$ such that $\lambda_{mi} \neq \lambda_{mj}$,\label{cond:lambda}
\item the columns of $W$ are constrained in a way that for all $i\neq j > d - d_1$,  the $i$th column cannot satisfy the constraints of the $j$th column as is nor after changing all signs in the $i$th column, and\label{cond:W}
\item there is at least one (strict) sign constraint in each of the last $d_1$ columns of $W$.\label{cond:sign}
\end{enumerate}
\end{proposition}
Condition~\ref{cond:sign} of Proposition~\ref{prop:ident1} fixes the signs in the last $d_1$ columns of $W$ and therefore the signs of the instantaneous effects of the corresponding shocks.  Changing the signs of the columns is effectively the same as changing the signs of the corresponding shocks, so Condition~\ref{cond:sign} is not restrictive, however (as the structural shock has a distribution that is symmetric about zero). The assumption that the identified shocks are the last $d_1$ shocks is neither restrictive as one may always reorder the structural shocks accordingly.  

For example, if $d=3$, $\lambda_{m1}\neq\lambda_{m3}$ for some $m$, and $\lambda_{m2}\neq\lambda_{m3}$ for some $m$, the third structural shock can be identified with the following constraints:
\begin{equation}\label{eq:exampleconstraints}
B_t=\begin{bmatrix}
* & * & *    \\
+ & + &  - \\
+ & + & + \\
\end{bmatrix}
\ \text{or} \
\begin{bmatrix}
- & * & + \\
- & + & -  \\
* & + & + \\
\end{bmatrix}
\ \text{or} \
\begin{bmatrix}
+ & + & 0  \\
* & * & *  \\
* & * & + \\
\end{bmatrix}
\end{equation} 
and so on, where "$*$" signifies that the element is not constrained, "$+$" denotes a strict positive and "$-$" a strict negative sign constraint, and "$0$" means that the element is constrained to zero. In the first example, Condition~\ref{cond:W} is satisfied because the last shock is assumed to move the last two variables to the opposite directions when the first two shocks are assumed to move them to the same direction, implying that the first two shocks cannot satisfy the constraint imposed on the last shock (as is nor after changing all signs of the impact responses). Similarly in the second example, the last shock moves to opposite directions the variables that the first two shocks move to the same direction. The last example imposes a zero constraint for the impact response of the first variable to the last shock, while the first two shocks impose strict sign constraints. Since the non-zero impact responses of the first two shocks cannot satisfy the zero constraint of the last shock, Condition~\ref{cond:W} is satisfied. By using sign and zero constraints in this manner, it is easy to produce further examples that lead to the identification of the last shock. 

Imposing sign or zero constraints on $W$ equals to placing them on $B_t$, so they can be justified economically. Under Assumption~\ref{as:eigenvalues}, the model is statistically identified prior to imposing the constraints, making the parameter constraints required in Condition~\ref{cond:W} also testable. This different to the conventional SVAR setup in which the identifying constraints cannot be validated statistically \cite[e.g.,][Chapters~8 and 10]{Kilian+Lutkepohl:2017}. Similarly to the conventional SVAR model, labelling the shocks formally with the economic shocks of interest, however, requires the identification constraints to be economically motivated. As Proposition~\ref{prop:ident1} shows and the examples in~(\ref{eq:exampleconstraints}) demonstrate, our method facilitates finding economically plausible identification constraints by flexibly using sign constraints alone or in combination with zero constraints. A point identification can be obtained even with only sign constraints, while in the conventional SVAR setup, sign constraints alone lead to a set identification only \cite[see e.g.,][Chapter~13]{Kilian+Lutkepohl:2017}. If Assumption~\ref{as:eigenvalues} fails, the structural GMVAR model is not fully identified and the problem of testing the parameter constraints is non-standard, which is briefly addressed in the next section.

\subsection{Identification of the shocks under partial identification of the model}\label{sec:partial_ident}
If Assumption~\ref{as:eigenvalues} is violated and the structural GMVAR model is thus not statistically identified, the shocks of interest can still be identified with Proposition~\ref{prop:ident1} if Condition~\ref{cond:lambda} is satisfied. When the shocks of interest do not satisfy Condition~\ref{cond:lambda}, their identification requires stronger constraints than in Proposition~\ref{prop:ident1}. Therefore, we present the following proposition that provides sufficient criteria for global identification of the last $d_1$ shocks when Condition~\ref{cond:lambda} fails; specifically, when exactly one of the eigenvalues $\lambda_{mi}$ with $i\neq j > d - d_1$ is identical to $\lambda_{mj}$ for all $m$. For simplicity, we assume that only one of the shocks with identical eigenvalues is to be identified, i.e., $i\leq d - d_1$ above.

\begin{proposition}\label{prop:ident2}
Let $d_1<d$. Consider the matrix decomposition of Proposition~\ref{prop:ident1} and further suppose that for $j=d-d_1+1$ and some $i\leq d-d_1$, we have $\lambda_{mi}=\lambda_{mj}$ for all $m$, but for all $l\centernot\in\lbrace i,j \rbrace$, $\lambda_{ml}\neq\lambda_{mj}$ for some $m$. Then, the last $d_1$ structural shocks are uniquely identified if Conditions~\ref{cond:lambda}-\ref{cond:sign} of Proposition~\ref{prop:ident1} are otherwise satisfied, and in addition
\begin{enumerate}[label={(\arabic*)}]
\setcounter{enumi}{3}
\item the column $i\leq d-d_1$ of $W$ such that $\lambda_{mi}=\lambda_{mj}$ for all $m$ has at least one (strict) sign constraint and the $j$th column has a zero constraint where the $i$th column has the (strict) sign constraint. \label{cond:zero}
\end{enumerate}
\end{proposition}
Note that the assumption $j=d-d_1+1$ is made without loss of generality, as the structural shocks can always be reordered accordingly by also reordering the columns of $W$ (including the constraints) and the eigenvalues $\lambda_{mi}$ correspondingly.

To exemplify, if $d=4$, $\lambda_{m1}\neq\lambda_{m4}$ for some $m$, $\lambda_{m2}\neq\lambda_{m4}$ for some $m$, and $\lambda_{m3}=\lambda_{m4}$ for all $m$, the following constraints lead to global identification of last shock:
\begin{equation}\label{eq:exampleconstraints2}
B_t=\begin{bmatrix}
* & * & * & *\\
* & * & + & 0\\
+ & + & * & -\\
+ & + & * & +\\
\end{bmatrix}
\ \text{or} \
\begin{bmatrix}
* & * & - & 0\\
* & * & * & *\\
+ & - & * & +\\
- & + & * & +\\
\end{bmatrix}
\ \text{or} \
\begin{bmatrix}
+ & - & - & 0\\
* & * & * & *\\
* & * & * & *\\
* & * & * & +\\
\end{bmatrix}
\end{equation} 
and so on. Condition~\ref{cond:zero} is satisfied in each of the above examples, because the third shock has a strict sign constraint for the variable that the last shock imposes a strict zero constraint. As is demonstrated above, the structural shocks can often be identified with flexible constraints even when some of the eigenvalues are identical for all regimes.
 
Under Conditions~\ref{cond:lambda} and \ref{cond:sign} of Proposition~\ref{prop:ident1}, the additional constraints on $W$ were stated testable because they are overidentifying and statistical identification of the model can always be achieved by fixing the ordering of the eigenvalues $\lambda_{mi}$, as long as none of the pairs of $\lambda_{mi}$, $i=1,...,d$, is identical for all $m=2,...,M$ (Assumption~\ref{as:eigenvalues}). In the setup of Proposition~\ref{prop:ident2}, however, when $\lambda_{mi}=\lambda_{mj}$ for all $m$ and some $i\neq j$, the model is not generally identified even when one fixes a unique ordering for the eigenvalues and the columns of $W$. Also, even if Condition~\ref{cond:zero} of Proposition~\ref{prop:ident2} is satisfied, only partial identification of the B-matrix is obtained since nothing guarantees unique identification of the $i$th column of $W$, which would require stronger conditions. Consequently, the model is not identified under the null nor the alternative hypothesis when testing for the constraints in Conditions~\ref{cond:W} and \ref{cond:zero}, making the testing problem non-standard and the conventional asymptotic distributions of the likelihood ratio and Wald test statistics unreliable. The same applies when one tests the equality of the eigenvalues in order to assess the validity of Condition~\ref{cond:lambda} of Proposition~\ref{prop:ident1}, as the model is not identified under the null. Deriving formal tests under no identification is, however, a major task and beyond the scope of this paper.\footnote{\cite{Lutkepohl+Meitz+Netsunajev+Saikkonen:2021} discussed a related testing problem under no identification and developed an asymptotic Wald type test for testing equality of the $\lambda_{mi}$ parameters in a linear SVAR model incorporating two volatility regimes with a known change point and (reduced form) shocks arriving from a class of elliptical distributions. \cite{Meitz+Saikkonen:2021}, on the other hand, studied the asymptotic properties of a likelihood ratio test statistic under no identification when testing for the number of regimes in mixture models with Gaussian conditional densities. One of the studied models is the GMAR model \citep{Kalliovirta+Meitz+Saikkonen:2015}, which is the univariate counterpart of the GMVAR model \citep{Kalliovirta+Meitz+Saikkonen:2016}.}

If more than two eigenvalues are identical for all $m=2,...,M$ but they are not all identical, it may still be possible to find flexible conditions for identification of the shocks. Specifically, the idea utilized in the proof of Proposition~\ref{prop:ident2} (presented in Appendix~\ref{sec:proofs}) can be applied to larger numbers of identical eigenvalues. If all the eigenvalues are identical for all covariance matrices, then $\Omega_m=\lambda_{m1}\Omega_1$ and the identification condition is the same as for the conventional SVAR model \citep[which is given, for example, in][Section 9.1.2 for the B-model]{Lutkepohl:2005}. As a general remark, observe that constraining an element of $B_t$ to be any constant other than zero is infeasible, because all elements on the right side of (\ref{eq:bt}) are either zero or time-varying due to the time-varying mixing weights.\footnote{We have focused on the B-model, where the structure is imposed on the contemporaneous relations of the shocks. Alternatively, one may consider the "A-model" setup in which the structure is placed on the contemporaneous relations of the observable variables governed by the "A-matrix" \citep[see, e.g.,][Section~9.1.1]{Lutkepohl:2005}. The A-model is obtained implicitly from the B-model (\ref{eq:sgmvar}) and (\ref{eq:sgmvarerr}) by defining the A-matrix as $A_t\equiv B_t^{-1}$, where $B_t$ is given by (\ref{eq:bt}). In this case, the structural model Equation~(\ref{eq:sgmvar}) becomes 
$$
A_ty_t = \sum_{m=1}^M(A_t\phi_{m,0}+\sum_{i=1}^pA_tA_{m,i}y_{t-i})+e_t,
$$
thereby incorporating continuously varying intercepts and coefficient matrices due to the constant variance normalization of the structural shocks. In practice, however, one needs to carefully derive how any specific constraint on $A_t$ can be imposed by restricting $W$.}

\subsection{Maximum likelihood estimation}\label{sec:estim}
The parameters of the reduced form GMVAR model are collected to the vector $\boldsymbol{\theta}=(\boldsymbol{\vartheta}_1,...,\boldsymbol{\vartheta}_M,\alpha_1,...,\alpha_{M-1})$  $((M(d^2p + d + d(d+1)/2) + 1) \times 1)$, where $\boldsymbol{\vartheta}_m=(\phi_{m,0},\text{vec}(A_{m,1}),....,\text{vec}(A_{m,p}),\text{vech}(\Omega_m))$, vec is a vectorization operator that stacks the columns of a matrix on top of each other, and vech stacks the columns of a matrix from the main diagonal downwards (including the main diagonal). The last mixing weight parameter $\alpha_M$ is omitted because it is obtained from the constraint $\sum_{m=1}^M \alpha_m=1$. 

Using the notation described in Section~\ref{sec:gmvar}, indexing the observed data as $y_{-p+1},...,y_0,y_1,...,y_T$, and assuming that the initial values $\boldsymbol{y}_0=(y_{-p+1},...,y_0)$ are stationary, the exact log-likelihood function of the reduced form GMVAR model takes the form \citep[Equations~(9) and (10)]{Kalliovirta+Meitz+Saikkonen:2016}
\begin{equation}\label{eq:loglik1}
L_t(\boldsymbol{\theta})=\log \left(\sum_{m=1}^M \alpha_m n_{dp}(\boldsymbol{y}_0;\boldsymbol{1}_p\otimes \mu_m, \boldsymbol{\Sigma}_m)   \right) + \sum_{t=1}^T l_t(\boldsymbol{\theta}),
\end{equation}
where
\begin{equation}\label{eq:loglik2}
l_t(\boldsymbol{\theta})=\log \left(\sum_{m=1}^M\alpha_{m,t} (2\pi)^{-d/2}\det(\Omega_m)^{-1/2}\exp\left\lbrace -\frac{1}{2}(y_t - \mu_{m,t})'\Omega_m^{-1}(y_t - \mu_{m,t}) \right\rbrace   \right).
\end{equation}
If it does not seem reasonable to assume that the initial values are stationary, one may condition on them and base the estimation on the conditional log-likelihood function, which is obtained by dropping the first term on the right side of (\ref{eq:loglik1}).

The reduced form GMVAR model can be estimated by maximizing the exact or conditional likelihood function in (\ref{eq:loglik1}) and (\ref{eq:loglik2}) with respect to the parameter $\boldsymbol{\theta}$. To ensure identification, the parameter space should be constrained so that the mixture components cannot be 'relabelled', for instance, by assuming that the mixing weight parameters are in a decreasing order,  $\alpha_M>\cdots\alpha_1>0$, and $\boldsymbol{\vartheta}_i=\boldsymbol{\vartheta}_j$ only if $i=j$ \citep[Equation~(11)]{Kalliovirta+Meitz+Saikkonen:2016}. If $M=2$, the structural GMVAR model is then obtained by simultaenously diagonalizing the reduced form error covariance matrices as discussed Section~\ref{sec:tworegime}. However, should overidentifying restrictions be imposed on $B_t$ through $W$ or if $M\geq 3$, it is more convenient to reparametrize the model with $W$ and $\Lambda_m$, $m=2,...,M$, instead of $\Omega_1,...,\Omega_M$ and maximize the log-likelihood function subject to the new set of parameters and constraints. In this case, the decomposition~(\ref{eq:decomp}) is plugged in to the log-likelihood function and the $\text{vech}(\Omega_1),...,\text{vech}(\Omega_M)$ are replaced with $\text{vec}(W),\lambda_2,...,\lambda_M$, where $\lambda_m=(\lambda_{m1},...,\lambda_{md})$, in the parameter vector $\boldsymbol{\theta}$.

Maximizing the complex and highly multimodal log-likelihood function can be challenging in practice,  particularly if there are more than two regimes. Following \cite{Dorsey+Mayer:1995}, \cite{Meitz+Preve+Saikkonen:2018,Meitz+Preve+Saikkonen:2021}, and \cite{Virolainen:2021,uGMAR}, we employ a two-phase estimation procedure where, in the first phase, a genetic algorithm is used to find starting values for a gradient based method which then, in the second phase, often converges to a nearby local maximum or saddle point. The genetic algorithm in the accompanying R package gmvarkit \citep{gmvarkitnormal} has been modified to improve its performance significantly, and it functions similarly to the one described in \cite{Virolainen:2021} for the univariate GMAR \citep{Kalliovirta+Meitz+Saikkonen:2015}, StMAR \citep{Meitz+Preve+Saikkonen:2021}, and G-StMAR \citep{Virolainen:2021} models. In order to obtain reliable results, a (sometimes very large) number of estimation rounds should be performed, for which gmvarkit makes use of parallel computing.

\section{Impulse response analysis}\label{sec:girf}
The expected effects of the structural shocks in the SGMVAR model generally depend on the initial values as well as on the sign and size of the shock, which makes the conventional way of calculating impulse responses unsuitable \citep[see, e.g.,][Chapter~4]{Kilian+Lutkepohl:2017}. Following \cite{Koop+Pesaran+Potter:1996} and \citet[Section~18.2.2]{Kilian+Lutkepohl:2017}, we therefore consider the generalized impulse response function (GIRF) defined as
\begin{equation}\label{eq:girf}
\text{GIRF}(h,\delta_j,\mathcal{F}_{t-1}) = \text{E}[y_{t+h}|\delta_j,\mathcal{F}_{t-1}] - \text{E}[y_{t+h}|\mathcal{F}_{t-1}],
\end{equation}
where $h$ is the chosen horizon and $\mathcal{F}_{t-1}=\sigma\lbrace y_{t-j},j>0\rbrace$ as before. The first term on the right side of (\ref{eq:girf}) is the expected realization of the process at time $t+h$ conditionally on a structural shock of size $\delta_j \in\mathbb{R}$ in the $j$th element at time $t$ and the previous observations. The latter term on the right side is the expected realization of the process conditionally on the previous observations only. The GIRF thus expresses the expected difference in the future outcomes when the structural shock of size $\delta_j$ in the $j$th element hits the system at time $t$ as opposed to all shocks being random. 

It is easy to see that the SGMVAR model has a $p$-step Markov property, so conditioning on (the $\sigma$-algebra generated by) the $p$ previous observations $\boldsymbol{y}_{t-1}=(y_{t-1},...,y_{t-p})$ is effectively the same as conditioning on $\mathcal{F}_{t-1}$ at time $t$ and later. The history $\boldsymbol{y}_{t-1}$ can be either fixed or random, but with random history the GIRF becomes a random vector, however. Using fixed $\boldsymbol{y}_{t-1}$ makes sense when one is interested in the effects of the shock at a particular point of time, whereas more general results are obtained by assuming that $\boldsymbol{y}_{t-1}$ follows the stationary distribution of the process. If one is, on the other hand, interested in a specific regime, $\boldsymbol{y}_{t-1}$ can be assumed to follow the stationary distribution of the corresponding component model. 

The GIRF and its distributional properties can be estimated with a Monte Carlo algorithm that generates (partial) realizations of the process and then takes the sample mean for point estimate. If $\boldsymbol{y}_{t-1}$ is random and follows the distribution $G$, the GIRF should be estimated for different values of $\boldsymbol{y}_{t-1}$ generated from $G$, and then the sample mean and sample quantiles can be taken to obtain the point estimate and confidence intervals that reflect the uncertainty about the initial value. Such an algorithm, adapted from \citet[pp. 135-136]{Koop+Pesaran+Potter:1996} and \citet[pp. 601-602]{Kilian+Lutkepohl:2017}, is given in Appendix~\ref{sec:montecarlo}.  

Because the SGMVAR model facilitates associating statistical characteristics and economic interpretations to the regimes, and because asymmetries in the GIRFs are caused by regime-switches, it may be of interest to also examine the effects of a structural shock to the mixing weights $\alpha_{m,t}$, $m=1,...,M$. We then consider the related GIRFs 
\begin{equation}
\text{GIRF}_{\alpha_m}(h,\delta_j,\mathcal{F}_{t-1}) = \text{E}[\alpha_{m,t+h}|\delta_j,\mathcal{F}_{t-1}] - \text{E}[\alpha_{m,t+h}|\mathcal{F}_{t-1}]
\end{equation}
for which point estimates and confidence intervals can be constructed similarly to (\ref{eq:girf}).

\section{Empirical application}\label{sec:empapp}
Our empirical application studies asymmetries in the effects of U.S. monetary policy shocks.  Asymmetric effects of U.S. monetary policy shocks have been studied, among others, by \cite{Weise:1999}, \cite{Garcia+Schaller:2002}, \cite{Lo+Piger:2005}, and \cite{Hoppner+Melzer+Neumann:2008},  who all found the effects of monetary policy shocks to production stronger during recessions (or low growth periods) than booms (or high growth periods). \cite{Weise:1999} also found evidence in favor of large and small shocks having different effects, and large positive and negative monetary shocks having different effects. \cite{Hoppner+Melzer+Neumann:2008} concluded that the real effects of monetary policy shocks have decreased over their sample period from 1962 to 2002. \cite{Tenreyro+Thwaites:2016}, on the other hand, found the effects of U.S. monetary policy shocks less powerful in recessions. %\cite{Peersman+Smets:2002} found that monetary policy shocks have larger effects on production during recessions than expansions in the Euro area, while \cite{Burgard+Neuenkirch+Nockel:2019} found the effects of contractionary monetary policy shocks stronger but less enduring during periods of "crisis" than "normal times".

We consider the quarterly U.S. data covering the period from 1954Q3 to 2021Q4 ($270$ observations) and consisting of four variables: real GDP, GDP implicit price deflator, producer price index (all commodities), and an interest rate variable. Our policy variable is the interest rate variable, which is the effective federal funds (FF) rate from 1954Q3 to 2008Q2. After that we replaced it with the \cite{Wu+Xia:2016} shadow rate, which is not constrained by the zero lower bound and also quantifies unconventional monetary policy measures.%\footnote{The Federal Open Market Committee targeted the FF rate between $0.00$ and $0.25$ from December 2008 to December 2015 (and more recently from March 2020 to March 2022). However, because the \cite{Wu+Xia:2016} shadow rate is somewhat higher than the FF rate in 2008Q4, but the FF rate and the shadow rate are close to each other in 2008Q2 and 2008Q3, we replaced the FF rate with the shadow rate already in 2008Q3. This produces a transition between the two interest rates such that it does not introduce any artificial jumps to the combined interest rate variable.} 
\footnote{The \cite{Wu+Xia:2016} shadow rate series was retrieved from the Federal Reserve Bank of Atlanta's website and the rest of the data were retrieved from the Federal Reserve Bank of St. Louis database.}

The GMVAR model requires stationary data, so the logarithms of the real GDP, GDP deflator, and producer price index need to be detrended. We detrend the logarithm of the real GDP by separating its cyclical component from the trend with the backward-looking Hodrick-Prescott (HP) filter and then considering the cyclical component.\footnote{The backward-looking HP filter was obtained from the two-sided HP filter by applying the filter up to horizon $t$, taking the last observation, and repeating this procedure for the full sample $t=1,...,T$.  In order to allow the series to start from any phase of the cycle, we applied the backward-looking filter to the full available sample from 1947Q1 to 2021Q4 before extracting our sample period from it. We computed the two-sided HP filter with the R package lpirfs \citep{lpirfs} by using the standard smoothing parameter value of $1600$. %For robustness, we also considered the first differences and the linear projection filter proposed by \cite{Hamilton:2018}. But they led to generalized impulse response functions where supposedly contractionary monetary policy shocks seemed to have significant, persistent expansionary effects on real GDP (not shown). Hence, we preferred the HP filter.
} It is thereby implicitly assumed that the monetary policy shock does not have permanent effects on real output. The logarithms of the price variables are detrended by taking the first difference and multiplying it by hundred, so the resulting series approximate the percentage growth rates. The interest rate variable is treated as stationary.

\begin{figure}[t]
    \centerline{\includegraphics[width=\textwidth - 2cm]{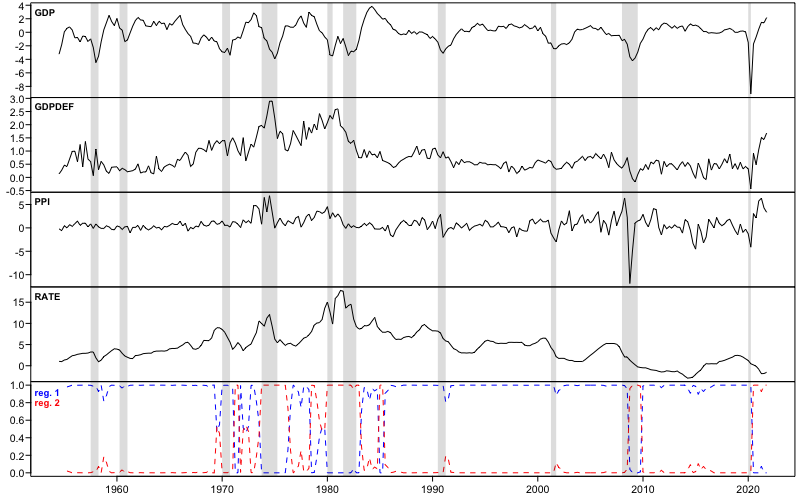}}
    \caption{Quarterly U.S. series covering the period from 1954Q3 to 2021Q4. The top panel presents the cyclical component of real GDP (GDP) which we separated from the trend using the one-sided Hodrick-Prescott filter. The second and third panels present the log-differences of GDP implicit price deflator (GDPDEF) and producer price index (PPI) multiplied by hundred. The fourth panel presents an interest rate variable, which is the effective federal funds from 1954Q3 to 2008Q2 and the \cite{Wu+Xia:2016} shadow rate from 2008Q3 to 2021Q4. The bottom panel shows the estimated mixing weights of the fitted GMVAR($3,2$) model. The shaded areas indicate the NBER based U.S. recessions.}
\label{fig:seriesplot}
\end{figure}

The series are presented in the first four top panels of Figure~\ref{fig:seriesplot} with the shaded areas indicating the periods of NBER based U.S. recessions. Throughout, we refer to the variables as GDP (output), GDPDEF (prices), PPI (commodity prices), and RATE (interest rate) without making it explicit that some of them are detrended. The (S)GMVAR model of autoregressive order $p$ and $M$ mixture components is referred to as (S)GMVAR($p,M$) model.

We select the order of our GMVAR model by first finding a suitable autoregressive order for a linear Gaussian VAR; that is, a GMVAR($p,1$) model. The AIC is minimized by the order $p=3$, suggesting that this might be the appropriate lag order for modelling autocorrelation. So we estimate a GMVAR($3,2$) model, which we find superior to the linear VAR. Graphical quantile residual diagnostics reveal that our GMVAR($3,2$) model adequately captures the autocorrelation structure of the series, but some of the conditional heteroskedasticity and excess kurtosis is not captured. In our view, the overall adequacy of the model is, nevertheless, reasonable enough for further analysis. Details on the model selection and quantile residual diagnostics are given in Appendix~\ref{sec:details}.  

The estimated mixing weights of the two regimes are presented in the bottom panel of Figure~\ref{fig:seriesplot}. The second regime (red) mainly dominates during periods of high inflation and interest rate in the 1970's and 1980's, after the collapse of Lehman Brothers in the Financial crisis until the end of 2009, and finally during the COVID-19 crisis from the third quarter of 2020 onwards. We refer to this regime as \textit{the unstable inflation regime}, as it generally exhibits high or volatile inflation. The first regime (blue) prevails when the second one does not: before 1970's, short periods during 1970's, and from the mid 1980's onwards but excluding the Financial crisis and the COVID-19 crisis (but including the first two quarters of 2020).  We refer to this regime as \textit{the stable inflation regime}, as it is characterized by moderate inflation. Details about the characteristics of the regimes are provided in Appendix~\ref{sec:details}.

\subsection{Identification of the monetary policy shock}\label{sec:identmone}
Decomposing the covariance matrices of the reduced form GMVAR($3,2$) model as in (\ref{eq:decomp}) gives the following estimates for the structural parameters:
\small
\begin{equation}\label{eq:estimates1}
\hat{W}=\begin{bmatrix*}
\phantom{-}\boldsymbol{0.14} \ (0.054) & \phantom{-}\boldsymbol{0.22} \ (0.065) & \boldsymbol{0.44} \ (0.053) &  -0.13 \ (0.136) \\
          \boldsymbol{-0.20} \ (0.014) &           -0.05 \ (0.028) & \boldsymbol{0.07} \ (0.012) &           -0.00 \ (0.022) \\
\phantom{-}0.00 \ (0.168) &           \boldsymbol{-1.03} \ (0.078) & \boldsymbol{0.47} \ (0.110) &           -0.06 \ (0.158) \\
\phantom{-}0.03 \ (0.029) & \phantom{-}0.03 \ (0.041) & 0.18 \ (0.100) & \phantom{-}\boldsymbol{0.30} \ (0.055) \\
\end{bmatrix*},
\ \
\hat{\lambda}_2 = \begin{bmatrix}
 \boldsymbol{\phantom{1}1.08} \ (0.227) \\
 \boldsymbol{\phantom{1}3.02} \ (0.636) \\
 \boldsymbol{11.05} \ (2.473) \\
 \boldsymbol{18.20} \ (3.601) \\
\end{bmatrix},
\end{equation}
\normalsize
where the ordering of the variables is $y_t=(\text{GDP}_t, \text{GDPDEF}_t, \text{PPI}_t, \text{RATE}_t)$, the estimates $\hat{\lambda}_{2i}$ are in an increasing order (which fixes an arbitrary ordering for the columns of $\hat{W}$), and approximate standard errors are given in parentheses next to the estimates. The estimates that deviate from zero by more than two times their approximate standard error are bolded. We proceed by assuming that all the $\lambda_{2i}$, $i=1,...,4$, are different to each other, i.e., that Assumption~\ref{as:eigenvalues} holds, which leads to statistical identification of the model. After identifying the monetary policy shock, the robustness of our identification with respect to this unjustified assumption is discussed.%\footnote{

Based on the estimates and their standard errors in (\ref{eq:estimates1}), the first shock moves GDP and inflation to the opposite directions, whereas the second shock moves GDP and commodity price inflation to the opposite directions. Since the instantaneous movements of the interest rate variable are insignificant, these two shocks do not seem plausible candidates for the monetary policy shock. The third shock moves the interest rate variable on impact more significantly than the first two shocks, but since production and both prices move significantly to the same direction, its characteristics appear similar to an aggregate demand shock and not a monetary policy shock. The last shock moves the interest rate variable significantly, while it also moves GDP, inflation, and commodity price inflation to the opposite direction, which is consistent with many of the standard the economic theories \cite[e.g.,][and the references therein]{Gali:2015}. The impact effects of GDP, inflation, and commodity price inflation are, however, statistically insignificant and the response of inflation is very weak. Nevertheless, among the four structural shocks obtained for the model, the characteristics of the last shock mostly resemble those of a monetary policy shock, so we deem it as the monetary policy shock.

Identifying the monetary policy shock formally by Proposition~\ref{prop:ident1} requires such constraints to be imposed on $W$ that it can be unambiguously distinguished from the other shocks. 
%Our identifying constraints are summarized in the following equation, where the monetary policy shock is ordered last:
%
%\begin{equation}\label{eq:Wimposed}
%W = 
%\begin{bmatrix}
%* & + & * & - \\
%- & * & + & 0 \\
%* & - & * & - \\
%* & * & * & + \\
%\end{bmatrix}.
%\end{equation}
%
We assume that the monetary policy shock moves the GDP and commodity price inflation to the opposite direction from the interest rate variable. In addition, we impose a zero constraint on the instantaneous movement of inflation, as the unrestricted estimated is very close to zero compared to the approximate standard error, and it allows us to avoid making restrictive assumptions about the first and third shocks. The Wald test produces the $p$-value $0.92$ for the zero constraint, so it is not rejected.

To distinguish the monetary policy shock from the other shocks, we assume that the first and third shocks move inflation at impact. This is not economically restrictive (since the responses can be very small) but it is a statistically reasonable assumption, as the Wald test rejects the hypotheses that the impact responses are zero (jointly or individually) with $p$-values less than $10^{-7}$. The Wald test produces the $p$-value $0.056$ for the hypothesis that the second shock does not move inflation at impact, so we cannot reject it. Therefore, we assume that the second shock moves the GDP and commodity price inflation to the opposite directions, as the corresponding estimates are large compared to their approximate standard errors, making the constraints statistically sensible. %In our view, they are also economically plausible, as a shock to the commodity prices increases (or decreases) the average marginal cost of production and can thereby be expected to decrease (or increase) output.

%The identification in %(\ref{eq:Wimposed}) 
%Section~\ref{sec:identmone} produced the following estimates for the GMVAR($3,2$) model:
%
%\small
%\begin{equation}\label{eq:estimates2}
%\hat{W}_2=\begin{bmatrix*}
%\phantom{-}\boldsymbol{0.14} \ (0.054) & \phantom{-}\boldsymbol{0.22} \ (0.065) & \boldsymbol{0.45} \ (0.037) &           -0.12 \ (0.065) \\
%          \boldsymbol{-0.20} \ (0.014) &           -0.05 \ (0.028) & \boldsymbol{0.07} \ (0.012) & \phantom{-}0\phantom{.00   (0.000)} \\
%\phantom{-}0.00 \ (0.168) &           \boldsymbol{-1.03} \ (0.078) & \boldsymbol{0.47} \ (0.109) & -0.05 \ (0.068) \\
%\phantom{-}0.03 \ (0.028) & \phantom{-}0.03 \ (0.041) & \boldsymbol{0.17} \ (0.044) & \phantom{-}\boldsymbol{0.30} \ (0.026) \\
%\end{bmatrix*},
%\ \
%\hat{\lambda}_2 = \begin{bmatrix}
% \boldsymbol{\phantom{1}1.08} \ (0.227) \\
% \boldsymbol{\phantom{1}3.02} \ (0.640) \\
%           \boldsymbol{11.05} \ (2.583) \\
%           \boldsymbol{18.20} \ (4.398) \\
%\end{bmatrix},
%\end{equation}
%\normalsize
%
%The estimates changed only slightly from the unrestricted ones in (\ref{eq:estimates1}), and the negative impact responses of output and commodity prices in the fourth column remain statistically insignificant. 

The estimates obtained for the structural parameters under the above-described identification changed only slightly from the unrestricted ones in (\ref{eq:estimates1}), and they are presented in Appendix~\ref{sec:structparestimates}. The estimates for $\lambda_{23}$ and $\lambda_{24}$ are somewhat close to each other relative to their standard errors, but due to our zero constraint on the inflation, Proposition~\ref{prop:ident2} identifies the monetary policy shock even if $\lambda_{23}=\lambda_{24}$ (or $\lambda_{21}=\lambda_{24}$). The monetary policy shock is identified also if additionally $\lambda_{2i}=\lambda_{2j}$ for any $i,j=1,2,3$, so our identification is not particularly sensitive to the validity of the unjustified Assumption~\ref{as:eigenvalues} (while the approximate standard errors and the Wald test results are invalid if the assumption fails).

\subsection{Generalized impulse response functions}\label{sec:girfresults}
Due to the endogenously determined regime-switching probabilities and the fact that we allow the regime to switch as a result of a shock, there are multiple types of possible asymmetries. The impulse responses can vary depending on the initial value as well as on the sign and size of the shock. We study the state-dependence of the (generalized) impulse response functions by drawing initial values from the stationary distribution of each regime separately. Then, we calculate the $90\%$ confidence intervals that reflect uncertainty about the initial value within the given regime as is described in Section~\ref{sec:girf} and Appendix~\ref{sec:montecarlo}. Asymmetries related to the sign and size of the shock are studied by estimating the GIRFs for positive (contractionary) and negative (expansionary) one-standard-error (small) and two-standard-error (large) shocks. After estimating the GIRFs, they are scaled so that the peak effect of the interest variable is $25$ basis points within the first four quarters, making the responses to shocks of different sign and size comparable.\footnote{The GIRFs are scaled based on the peak response instead of the initial response, because the peak response is much higher compared to the initial response in the first regime than in the second regime. Scaling the GIRFs based on the initial response would then shift the response of the interest rate variable significantly higher in the first regime than in the second regime.}

\begin{figure}
    \centerline{\includegraphics[width=\textwidth - 2cm]{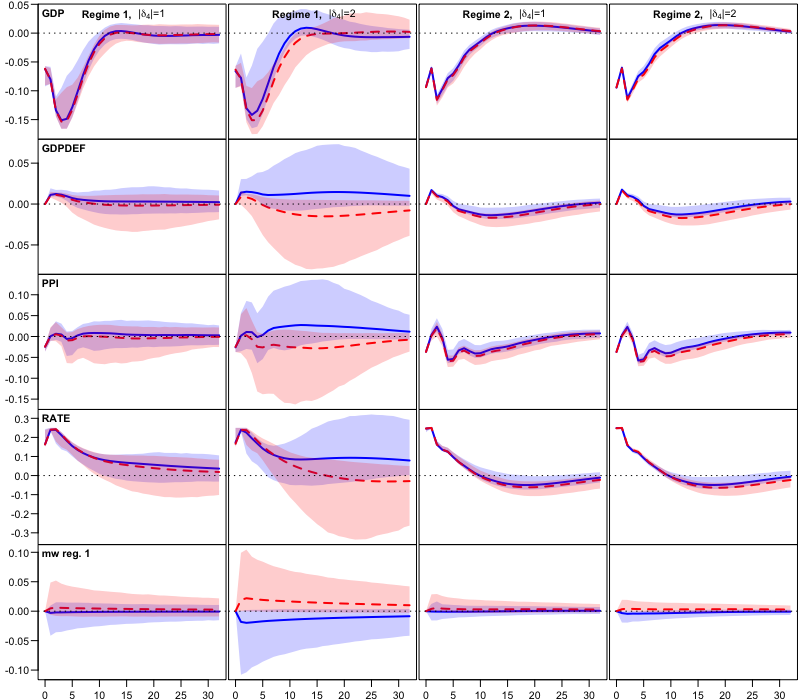}}
    \caption{Generalized impulse response functions $h=0,1,....,32$ quarters ahead estimated for the monetary policy shock identified in Section~\ref{sec:identmone} using $R_1=R_2=2500$ in the Monte Carlo algorithm presented in Appendix~\ref{sec:montecarlo}. From the top to bottom, the responses of production, prices, commodity prices, the interest rate variable, and the first regime's mixing weights are depicted in each row, respectively. The GIRFs are not accumulated for the prices variables, i.e., to responses are for inflation rates. The first [third] column shows the responses to a positive (blue solid line) and negative (red dashed line) one-standard-error shocks with the initial values generated from the stationary distribution of the first [second] regime. The second [fourth] column shows the responses to a positive and negative two-standard-error shocks with the initial values generated from the first [second] regime. All GIRFs have been scaled so that peak effect of the interest rate variable is $25$ basis points during the first four quarters. The shaded areas represent the $90\%$ confidence intervals that reflect uncertainty about the initial state within the given regime. Responses of the second regime's mixing weights are omitted because they are the negative of the first regime's mixing weights' responses.}
\label{fig:girfplot}
\end{figure}

Figure~\ref{fig:girfplot} presents the GIRFs $h=0,1,...,32$ quarters ahead estimated for the identified monetary policy shock.\footnote{We use $R_1=R_2=2500$ in the Monte Carlo algorithm, i.e., for each regime, size, and sign of the shock we draw $2500$ initial values, and for each of those initial values the GIRF is estimated based on $2500$ different sample paths.} %using the Monte Carlo algorithm presented in Appendix~\ref{sec:montecarlo} with $R_1=R_2=2500$, i.e., for each regime, size, and sign of the shock we draw $2500$ initial values, and for each of those initial values the GIRF is estimated based on $2500$ different sample paths. 
The GIRFs of inflation rate and commodity price inflation rate are not accumulated to levels. From top to bottom, the responses of GDP, inflation rate, commodity price inflation rate, interest rate, and the first regime's mixing weights are depicted in each row, respectively. The first [third] column shows the responses to small contractionary (blue solid line) and expansionary (red dashed line) shocks with the initial values generated from the stationary distribution of the first [second] regime. The second [fourth] column shows the responses to large contractionary and expansionary shocks with the initial values generated from the first [second] regime. The shaded areas are the $90\%$ confidence intervals that reflect uncertainty about the initial value within the given regime. The responses of the second regime's mixing weights are not depicted because they are the negative of those of the first regime.

In the first regime (the stable inflation regime; the first and second columns of Figure~\ref{fig:girfplot}), contractionary (expansionary) monetary policy shock causes a significant contraction (expansion) in the GDP, with the peak effect occurring after three to four quarters. On average, the response is hump shaped and decays to zero roughly after three years from impact.\footnote{By zero, we mean the expected observation if all the shocks were random. Accordingly, by positive we mean expected observations larger than that and by negative expected observations smaller than that.} As the confidence bounds show, the response switches sign from some of the starting values, while some of the starting values display persistent contraction. Particularly large shocks seem to cause a delayed expansion (contraction) from many of the starting values - shortly after the impact when the shock is contractionary and later when the shock is expansionary.  

The prices mostly rise in response to a contractionary monetary policy shock, although one would often expect a contractionary monetary policy shock to decrease inflation due to the decreased aggregate demand. This is often referred to as the price puzzle, and it has been discussed recently, for instance, in \citet[Section~3.3.2, and the references therein]{Ramey:2016}.\footnote{A popular explanation is that the Fed uses more information in predicting the future inflation than the autoregressive system of the variables included in the model \citep{Sims:1992}. Consequently, the identified monetary policy shock also contains a component that incorporates the Fed's endogenous response to the prediction of the future inflation that is not captured by the autoregressive system of the included variables. If the endogenous response is not strong enough to offset the predicted inflation, the impulse responses may then display a rise in the inflation. Another explanation proposes that the prices increase due to the cost-push effect of the monetary policy shock. An increase in the nominal interest rate increases the marginal cost of production of the firms who operate on borrowed money, and thereby decreases the aggregate supply and increases the price level \cite[e.g.,][]{Barth+Ramey:2001, Ravenna+Walsh:2006}. Several authors have, however, argued that the cost-channel is not likely strong enough to cause a price puzzle even in the short-run \cite[e.g.,][]{Rabanal:2007, Kaufmann+Scharler:2009, Castelnuovo:2012}. Nonetheless, we find our empirical results interesting, as the long-run price puzzle arises only from some of the starting values, while its occurrence is also sensitive to the sign and size of the shock. Moreover, as is discussed in Appendix~\ref{sec:modelselection}, enforcing linearity to the autoregressive dynamics makes the price puzzle worse.}
As is explained in Appendix~\ref{sec:indgirfres}, the prices rise because the shock drives the economy towards the unstable inflation regime, which has higher long-run inflation. 

On average, inflation does not move much in response to a small expansionary monetary policy shock, while the interest rate stays low relatively persistently and is accompanied with a roughly three years long expansion of the GDP. A large expansionary shock drives the economy towards the unstable inflation regime relatively more than a small expansionary shock, as the responses of the first regime's mixing weights show. Consequently, the inflation mostly increases and the interest rate variable increases relatively fast towards zero, and as the confidence bounds show, from many of the starting values the interest rate variable overshoots significantly. It is shown in Appendix~\ref{sec:indgirfres} that these GIRFs are the ones that also display particularly high peak inflation, and that the significant monetary policy tightening is accompanied with a persistent contraction of the GDP after the initial expansion.

In the second regime (the unstable inflation regime; the third and fourth columns of Figure~\ref{fig:girfplot}), contractionary (expansionary) monetary policy shock causes a strong contraction (expansion) of the GDP, with the peak effect occurring after two quarters. After roughly three years, from most of the starting values the response overshoots and becomes expansionary (contractionary) before decaying to zero. Inflation rate and commodity price inflation rate rise after the impact and then decrease significantly for several years before the price levels stabilize. 

The interest rate decays towards zero for roughly two years after which, on average, it decreases (increases) slightly below (above) zero before returning to zero. The response of the interest rate switches its sign less significantly when the shock is contractionary, when also the inflationary effects of the shock are slightly weaker. %It therefore seems that the interest rate possibly overshoots as an endogenous response to stabilize the price level, possibly causing the delayed expansion (contraction) of the GDP before it decays to zero. 
The scaled GIRFs are almost identical for small and large shocks, so there does not appear to be much asymmetries with respect to the size of the shock. The asymmetries are weak, because the monetary policy shock has relatively weak effect on the regime-switching probabilities, as the scaled responses of the first regime's mixing weights show (the third and fourth bottom panels of Figure~\ref{fig:girfplot}).

Overall, expansionary and contractionary monetary policy shocks seem to both mostly increase the probability of the unstable inflation regime, significantly more so if the economy is in the stable inflation regime when the shock arrives (the bottom panels in Figure~\ref{fig:girfplot}). In the stable inflation regime, a large shock increases the probability of entering the unstable inflation regime relatively more than a small shock, and often propagates high and persistent inflation, which is followed by a significant monetary policy tightening and a persistent contraction of the GDP. On average, the real effects of the monetary policy shock are somewhat stronger in the stable inflation regime than in the unstable inflation regime, but the (average) effects die out equally fast.

\section{Summary}\label{sec:summary}
We introduced a structural version of the Gaussian mixture vector autoregressive model \citep{Kalliovirta+Meitz+Saikkonen:2016} that incorporates endogenously determined mixing weights and a time-varying B-matrix. We showed that our model generally identifies the structural shocks up to ordering and sign, but does not reveal which column of the B-matrix is related to which shock. Since the B-matrix is also subject to an estimation error, we made use of the matrix decomposition proposed by \cite{Lanne+Lutkepohl:2010} and \cite{Lanne+Lutkepohl+Maciejowska:2010} and derived general conditions for formally identifying any subset of the shocks. This led to flexible identification conditions, and some of the constraints are also testable. For impulse response analysis, we utilized the generalized impulse response function \citep{Koop+Pesaran+Potter:1996} and proposed a Monte Carlo algorithm for its estimation by making use of the known stationary distribution of the SGMVAR process. The paper is accompanied with the CRAN distributed R package gmvarkit \citep{gmvarkitnormal}, which provides a comprehensive set of tools for numerical analysis of the model. 

Our empirical application studied asymmetries in the expected effects of monetary policy shocks in the U.S.  using a quarterly series covering the period from 1954Q3 to 2021Q4. Our SGMVAR model identified two regimes: a stable inflation regime and an unstable inflation regime. The unstable inflation regime is characterized by high or volatile inflation, and it mainly prevails in the 1970's, early 1980's, during the Financial crisis, and in the COVID-19 crisis from 2020Q3 onwards. The stable inflation regime, in turn, is characterized by moderate inflation, and it prevails when the stable inflation regime does not. We found the effects of the monetary policy shock relatively symmetric in the unstable inflation regime, as it rarely causes a switch to the stable inflation regime. A contractionary (expansionary) monetary policy shock appears to first increase (decrease) inflation after which the inflation significantly decreases (increases) for several years. The strong contraction (expansion) in the cyclical component of GDP lasts for roughly three years and is followed by a relatively mild expansion (contraction) along with the interest rate variable overshooting to the negative (positive) side.

The effects of the monetary policy shock were found strongly asymmetric in the stable inflation regime with respect to the initial state of the economy as well as to the sign and size of the shock. A large shock often causes relatively stronger inflationary effects than a small shock, while both contractionary and expansionary shocks seem to increase inflation by driving the economy towards the unstable inflation regime. A small expansionary shock does not move prices much on average, but a large expansionary shock often drives the economy towards the unstable inflation regime and propagates high and persistent inflation. The high inflation is followed by a significant monetary policy tightening and a persistent contraction of the GDP after the initial expansion. On average, the real effects of the monetary policy shock were found somewhat stronger in the stable inflation regime than in the unstable inflation regime.

%\section*{Acknowledgements}
%This work was supported by the Academy of Finland under Grant 308628. The author thanks Markku Lanne, Mika Meitz,  and Pentti Saikkonen who helped to improve this paper substantially. The author also thanks Henri Nyberg and Antti Ripatti for the useful comments.

%\section*{Declaration of interest statement}
%The author has no conflict of interest to declare.

\bibliography{refs}

\pagebreak
\begin{appendices}

\section{Proofs}\label{sec:proofs}

\subsection{Proof of Lemma~\ref{lemma1}}
Consider $M$ positive definite $(d\times d)$ covariance matrices $\Omega_m$, $m=1,...,M$ and suppose $B$ is any invertible $(d\times d)$ matrix such that $B^{-1}\Omega_m B'^{-1}$  are diagonal matrices with strictly positive diagonal entries. It follows that $B^{-1}\Omega_1 B'^{-1}=\Lambda_m^{-1}B^{-1}\Omega_m B'^{-1}$ for some $(d\times d)$ diagonal matrices $\Lambda_m^{-1}$, $m=2,...,M$, that have strictly positive diagonal entries. Elementary matrix algebra then shows that these identities are equivalent to $B\Lambda_m  = \Omega_m\Omega_1^{-1}B$, $m=2,...,M$.
%\begin{align*}
 %B^{-1}\Omega_1 B'^{-1} &= \Lambda_m^{-1}B^{-1}\Omega_m B'^{-1}\\
%\iff  B^{-1}\Omega_1  &= \Lambda_m^{-1}B^{-1}\Omega_m \\
%\iff  B^{-1}  &= \Lambda_m^{-1}B^{-1}\Omega_m\Omega_1^{-1} \\
%\iff  \Lambda_mB^{-1}  &= B^{-1}\Omega_m\Omega_1^{-1} \\
%\iff  B\Lambda_mB^{-1}  &= \Omega_m\Omega_1^{-1} \\
%\iff  B\Lambda_m  &= \Omega_m\Omega_1^{-1}B \\
%\end{align*}
Thus, the matrices $B$ and $\Lambda_m$ solve the eigenvalue problem of $\Omega_m\Omega_1^{-1}$ with the diagonal of $\Lambda_m=\text{diag}(\lambda_{m1},...,\lambda_{md})$ containing the strictly positive eigenvalues and the columns of $B$ being the related eigenvectors. Since this holds for any invertible $(d\times d)$ matrix $B$ that simultaneously diagonalizes the covariance matrices, it is also a necessary property of a time-varying B-matrix $B_t$.

Suppose also $BA$ solves the eigenvalue problems of $\Omega_m\Omega_1^{-1}$, $m=2,...,M$, for some invertible $(d\times d)$ matrix $A$. That is, $BA\Lambda_m = \Omega_m\Omega_1^{-1}BA$ which is equivalent to $A\Lambda_mA^{-1} = B^{-1}\Omega_m\Omega_1^{-1}B$. But since $B^{-1}\Omega_m\Omega_1^{-1}B=\Lambda_m$, this implies that $A\Lambda_mA^{-1} = \Lambda_m$, which is equivalent to  $A\Lambda_m = \Lambda_mA$. Thus, $\lambda_{mi}a_{ij}=\lambda_{mj}a_{ij}$ where $a_{ij}$ is the $ij$th element of $A$. It follows that $a_{ij}=0$ if $\lambda_{mi}\neq\lambda_{mj}$ for some $m$, implying that $A$ is diagonal matrix under Assumption~\ref{as:eigenvalues}, and $BA$ multiplies each of the columns of $B$ by a scalar. It is well known that eigenvalues of a matrix are unique (up to order), but since the diagonal elements of $\Lambda_m$ can be in any order, so can the related eigenvectors that are the columns of $B$. That is,  $B$ is unique up to scalar multiples and ordering of its columns. Since the above holds for any appropriate B-matrices $B$ and $BA$, it holds also for a time-varying B-matrix $B_t$ at each $t$.$\blacksquare$ % Since the eigenvectors in the columns of B are linearly independent, any nonzero (as A is invertible) vector can be formed by the linear combinations in BA by choosing A accordingly. Thus, it suffices to the study the matrices BA. 

\subsection{Proof of Proposition~\ref{prop:Bmat_uniqueness}}
Lemma~\ref{lemma1} shows that the B-matrix $B_t$ is unique up to scalar multiples and reordering of its columns. Suppose the conditional covariance matrix of the structural error is normalized to a constant diagonal matrix with strictly positive diagonal entries, say $C$. That is, $\sum_{m=1}^M\alpha_{m,t}B_t^{-1}\Omega_mB_t'^{-1}=C$, which is equivalent to $\sum_{m=1}^M\alpha_{m,t}\Omega_m=B_tCB_t'$.  Suppose that this identity also holds with another B-matrix, $B_tE_t$, where $E_t$ is a possibly time-varying, invertible $(d\times d)$ matrix. We have $\sum_{m=1}^M\alpha_{m,t}(B_tE_t)^{-1}\Omega_m(B_tE_t)'^{-1}=C$, which is equivalent to $\sum_{m=1}^M\alpha_{m,t}\Omega_m=(B_tE_t)C(B_tE_t)'$.  Thus, $B_tCB_t'=(B_tE_t)C(B_tE_t)'$. By Lemma~\ref{lemma1}, the B-matrix is unique up to scalar multiples and reordering of its columns, so with a given ordering of the columns, $E_t$ is a diagonal matrix. It then follows from $B_tCB_t'=(B_tE_t)C(B_tE_t)'$ that $C=E_tCE_t$, which in turn implies $c_{i}=e_{t,i}^2c_{i}$, where $c_{i}$ and $e_{t,i}$ are the $i$th diagonal elements of $C$ and $E_t$, respectively.  Therefore, $e_{t,i}=\pm 1$, implying that with a given ordering of the columns, (for each $t$) $B_t$ is unique up to changing all signs in a column. Therefore, $B_t$ is unique up ordering of its columns and changing all signs in a column.$\blacksquare$  

\subsection{Proof of Proposition~\ref{prop:ident1}}
Let $\Omega_1,...,\Omega_M$ be positive definite covariance matrices. We consider the decomposition $\Omega_1=WW'$ and $\Omega_m=W\Lambda_mW', \ \ m=2,...,M,$ where $\Lambda_m=\text{diag}(\lambda_{m1},...,\lambda_{md})$, $\lambda_{mi}>0$ ($i=1,..,d$), contains the eigenvalues of $\Omega_m\Omega_1^{-1}$ in the diagonal and the columns of the nonsingular $W$ are the related eigenvectors. The decomposition always exists when $M=2$ \citep[see, e.g.,][Theorem~A9.9]{Muirhead:1982} but not necessarily when $M\geq 3$. In the following, we assume the covariance matrices satisfy the decomposition.

Repeating some of the proof in \citet[p. 130; see also the proof of Theorem~A9.9 in \citealp{Muirhead:1982}]{Lanne+Lutkepohl+Maciejowska:2010} for convenience, suppose that we also have $\Omega_1=DD'$ and $\Omega_m=D\Lambda_m D'$, $m=2,...,M$, for some nonsingular $(d\times d)$ matrix $D$. Because $D^{-1}WW'D'^{-1}=D^{-1}\Omega_1D'^{-1}=I_d$, the matrix $Q'\equiv D^{-1}W$ is orthogonal, and hence, $D=WQ$ and $\Lambda_m Q = Q\Lambda_m$. %Then $\Lambda=W^{-1}\Omega_2W'^{-1}=W^{-1}M\Lambda M'W'^{-1}=Q\Lambda Q'$ so $\Lambda Q = Q\Lambda$.
It follows that $\lambda_{mi}q_{ij}=\lambda_{mj}q_{ij}$ where $q_{ij}$ is the $ij$th element of $Q$. Thus, $q_{ij}=0$ if $\lambda_{mi}\neq\lambda_{mj}$ for some $m$. Assuming that this (Condition~\ref{cond:lambda}) is satisfied by the last $d_1\in \lbrace 1,...,d \rbrace$ eigenvalues, it follows that $Q$ is a block-diagonal matrix with two blocks in the diagonal. Denoting $d_0 \equiv d-d_1$, the first block is a $(d_0 \times d_0)$ matrix and the second one is a $(d_1\times d_1)$ diagonal matrix with $q_{d_0+1,d_0+1},...,q_{d,d}$ in the diagonal (if $d_1=d$, $Q$ simply reduces to a diagonal matrix).

As the blocks in the diagonal of an orthogonal block-diagonal matrix are orthogonal and the real eigenvalues of a diagonal orthogonal matrix are $\pm 1$, it follows that the real eigenvalues of the second block in the diagonal of $Q$ are $\pm 1$. Then, because the eigenvalues of a block-diagonal matrix are the eigenvalues of the blocks in the diagonal, and eigenvalues of a diagonal matrix are its diagonal elements (and $Q$ is real), it must be that $q_{d_0+1,d_0+1},...,q_{d,d}$ are $\pm 1$. 

Thus, because $D=WQ$, the last $d_1$ columns of $W$ are unique up to changing all signs in a column for given $\Lambda_m$, $m=2,...,M$. Since $\Lambda_m$ are unique up to ordering of the diagonal elements and Condition~\ref{cond:W} fixes a unique ordering for the last $d_1$ columns of $W$ and hence also for the related eigenvalues $\lambda_{mi},i> d_0$, the last $d_1$ columns of the B-matrix~(\ref{eq:bt}) are uniquely identified up to changing all signs in a column. Finally, Condition~\ref{cond:sign} fixes the signs in the last $d_1$ columns of $W$ and consequently of $B_t$, implying that the last $d_1$ columns of the B-matrix are (globally) unique for given mixing weights $\alpha_{1,t},...,\alpha_{M,t}$. Moreover, if $d_1=d$, the decomposition~(\ref{eq:decomp}) of $\Omega_1,...,\Omega_M$ is (globally) unique.$\blacksquare$  

\section{Proof of Proposition~\ref{prop:ident2}}

Consider the matrix decomposition of $\Omega_m$, $m=1,...,M$, of Proposition~\ref{prop:ident1}. It is shown in the proof of Proposition~\ref{prop:ident1} that any $(d\times d)$ matrix $D$ that also satisfies $\Omega_1=DD'$ and $\Omega_m=D\Lambda_m D'$, $m=2,...,M$, can be presented as $D=WQ$ where $Q$ is orthogonal and $q_{ij}=0$ when $\lambda_{mi}\neq \lambda_{mj}$ for some $m$. Then, observe that the $j$th column of $WQ$ is a linear combination of the columns of $W$, with the multiplier of the $i$th column given by $q_{ij}$. Denoting $d_0\equiv d - d_1$, it follows that if $\lambda_{mi}=\lambda_{mj}$ for $i\neq j > d_0$ and all $m$, but for all $l\centernot\in\lbrace i,j \rbrace$, $\lambda_{ml}\neq\lambda_{mj}$ for some $m$, the $j$th column of $WQ$ is a linear combination of the $i$th and $j$th columns of $W$. But if the $j$th column (of $W$ and $WQ$) obeys a zero constraint where the $i$th column obeys a strict sign constraint (Condition~\ref{cond:zero}), the multiplier $q_{ij}$ must be zero. That is, under the conditions of Proposition~\ref{prop:ident2}, with $j=d_0+1$ and $i<j$, we have $q_{l,d_0+1}=0$ for all $l\neq d_0+1$ and $q_{lk}=0$ for all $l,k=d_0+2,..,d$ such that $l\neq k$.

By the above discussion, when $d_1>1$, $Q$ is a block-diagonal matrix with two blocks in the diagonal: the first one being a $(d_0+1\times d_0+1)$ matrix
\begin{equation}\label{eq:Qtilde}
\tilde{Q}\equiv
\begin{bmatrix}
q_{1,1} & \cdots & q_{1,d_0} & 0 \\
\vdots  & \ddots & \vdots & \vdots \\
q_{d_0,1} & \cdots & q_{d_0,d_0} & 0 \\
q_{d_0+1,1} & \cdots & q_{d_0+1,d_0} & q_{d_0+1,d_0+1} \\
\end{bmatrix}
\end{equation}
and the second one a $(d_1-1\times d_1-1)$ diagonal matrix with $q_{d_0+2,d_0+2},...,q_{d,d}$ in the diagonal. When $d_1=1$, we simply have $Q=\tilde{Q}$ where $\tilde{Q}$ is as in (\ref{eq:Qtilde}). Consequently, for $k>d_0$ the $k$th column of $WQ$ equals to the $k$th column of $W$ multiplied by $q_{k,k}$. It then remains to show that $q_{k,k}=\pm 1$ for all $k=d_0+1,...,d$, after which global uniqueness of the last $d_1$ columns of the B-matrix can be concluded with arguments similar to the proof of Proposition~\ref{prop:ident1}. 

Because only the last element of the last column of $\tilde{Q}$ is nonzero, the minors of the elements $q_{d_0+1,1}, ...., q_{d_0+1,d_0}$ are singular. Therefore, it follows from the cofactor presentation of the inverse of $\tilde{Q}$ \citep[e.g.,][Appendices~A4 and A5]{Muirhead:1982} that only the last element in the last column of the inverse of $\tilde{Q}$ is nonzero. Since $\tilde{Q}$ is orthogonal, as it is the upper-left block of the block-diagonal orthogonal matrix $Q$, %and blocks of orthogonal block diagonal matrix are also orthogonal, 
its transpose is also its inverse. Hence, only the last element in the last column of the transpose of $\tilde{Q}$ is nonzero. Also, by the definition of $\tilde{Q}$, only the last element in last row of the transpose of $\tilde{Q}$ is nonzero. That is, the transpose is of the form 
\begin{equation}\label{eq:Qtildetrans}
\tilde{Q}'=
\begin{bmatrix}
q_{1,1} & \cdots & q_{d_0,1} & 0 \\
\vdots  & \ddots & \vdots & \vdots \\
q_{1,d_0} & \cdots & q_{d_0,d_0} & 0 \\
0 & \cdots & 0 & q_{d_0+1,d_0+1} \\
\end{bmatrix},
\end{equation}
implying that
\begin{equation}
\tilde{Q}=
\begin{bmatrix}
q_{1,1} & \cdots & q_{1,d_0} & 0 \\
\vdots  & \ddots & \vdots & \vdots \\
q_{d_0,1} & \cdots & q_{d_0,d_0} & 0 \\
0 & \cdots & 0 & q_{d_0+1,d_0+1} \\
\end{bmatrix}.
\end{equation}
The matrix $Q$ is therefore an orthogonal block-diagonal matrix with two blocks in the diagonal. The first block is the upper-left $(d_0\times d_0)$ submatrix of $\tilde{Q}$ and the second block is the $(d_1\times d_1)$ diagonal matrix with $q_{d_0+1,d_0+1},...,q_{d,d}$ in the diagonal. Now reasoning similar to the proof of Proposition~\ref{prop:ident1} shows that $q_{k,k}=\pm 1$ for $k=d_0+1,...,d$.$\blacksquare$  

\section{Monte Carlo algorithm}\label{sec:montecarlo}
We present a Monte Carlo algorithm that produces point estimates and with random initial value $\boldsymbol{y}_{t-1}=(y_{t-1},...,y_{t-p})$ confidence intervals for the generalized impulse response function defined in (\ref{eq:girf}). Our algorithm is adapted from \citet[pp. 135-136]{Koop+Pesaran+Potter:1996} and \citet[pp. 601-602]{Kilian+Lutkepohl:2017}. We assume that the history $\boldsymbol{y}_{t-1}$ follows a known distribution $G$, which may be such that it produces a single outcome with probability one (corresponding to a fixed $\boldsymbol{y}_{t-1}$), or it can be the stationary distribution of the process or of a specific regime. In the following, $y_{t+h}^{(i)}(\delta_j,\boldsymbol{y}_{t-1})$ denotes a realization of the process at time $t+h$ conditional on the structural shock of magnitude $\delta_j$ in the $j$th element of $e_t$ hitting the system at time $t$ and on the $p$ observations $\boldsymbol{y}_{t-1}=(y_{t-1},...,y_{t-p})$ preceding the time $t$, whereas $y_{t+h}^{(i)}(\boldsymbol{y}_{t-1})$ denotes an alternative realization conditional on the history $\boldsymbol{y}_{t-1}$ only.

The algorithm proceeds with the following steps.
\begin{enumerate}\addtocounter{enumi}{-1}
\item Decide the horizon $H$, the numbers of repetitions $R_1$ and $R_2$, and the magnitude $\delta_j$ for the $j$th structural shock that is of interest. 

\item Draw an initial value $\boldsymbol{y}_{t-1}$ from $G$.\label{step1}  

\item Draw $H+1$ independent realizations of a shock $\varepsilon_t$ from $N(0,I_d)$. Also, draw an initial regime $m\in \lbrace 1,...,M \rbrace$ according to the probabilities given by the mixing weights $\alpha_{1,t},...,\alpha_{M,t}$ and compute the reduced form shock $u_t=W\Lambda_m^{1/2}\varepsilon_t$, where $\Lambda_1=I_d$. Then, compute the structural shock $e_t = B_t^{-1}u_t$ and impose the size $\delta_j$ on its $j$th element to obtain $e_t^*$. Finally, calculate the modified reduced form shock $u_t^*=B_te_t^*$.\footnote{The independent standard normal shocks $\varepsilon_t$ are introduced here to control random variation across the two sample paths $y_{t+n}^{(i)}(\delta_j,\boldsymbol{y}_{t-1})$ and $y_{t+n}^{(i)}(\boldsymbol{y}_{t-1})$.}\label{step2}

\item Use the modified reduced form shock $u_t^*$ and the rest $H$ standard normal shocks $\varepsilon_t$ obtained from Step~\ref{step2} to compute realizations $y_{t+h}^{(i)}(\delta_j,\boldsymbol{y}_{t-1})$ for $h=0,1,...,H$, iterating forward so that in each iteration the regime $m$ that generates the observation is first drawn according to the probabilities given by the mixing weights. At $h=0$, the initial regime and the modified reduced form shock $u_t^*$ calculated from the structural shock in Step~\ref{step2} is used. From $h=1$ onwards, the $h+1$th standard normal shock $\varepsilon_t$ is used to calculate the reduced form shock $u_{t+h}=W\Lambda_m^{1/2}\varepsilon_{t+h}$, where $\Lambda_1=I_d$ and $m$ is the selected regime.

\item  Use the reduced form shock $u_t$ and the rest $H$ the standard normal shocks $\varepsilon_t$ obtained from  Step~\ref{step2} to compute realizations $y_{t+h}^{(i)}(\boldsymbol{y}_{t-1})$ for $h=0,1,...,H$, so that the reduced form shock $u_t$ (calculated in Step~\ref{step2}) is used to compute the time $h=0$ realization. Otherwise proceed similarly to the previous step.

\item Calculate $y_{t+h}^{(i)}(\delta_j,\boldsymbol{y}_{t-1}) - y_{t+h}^{(i)}(\boldsymbol{y}_{t-1})$.\label{step5}

\item Repeat Steps~\ref{step2}-\ref{step5} $R_1$ times and calculate the sample mean of $y_{t+h}^{(i)}(\delta_j,\boldsymbol{y}_{t-1}) - y_{t+n}^{(i)}(\boldsymbol{y}_{t-1})$ for $h=0,1,...,H$ to obtain an estimate of the GIRF$(h,\delta_j,\boldsymbol{y}_{t-1})$.\label{step6}

\item Repeat Steps~\ref{step1}-\ref{step6} $R_2$ times to obtain estimates of GIRF$(h,\delta_j,\boldsymbol{y}_{t-1})$ with different starting values $\boldsymbol{y}_{t-1}$ generated from the distribution $G$. Then, take the sample mean and sample quantiles over the estimates to obtain point estimate and confidence intervals for the GIRF with random initial value.\label{step7}
\end{enumerate}
Notice that if a fixed initial value $\boldsymbol{y}_{t-1}$ is used, Step~\ref{step7} is redundant.

\section{Details on the empirical application}\label{sec:details}

\subsection{Model selection}\label{sec:modelselection}
The maximum likelihood (ML) estimation of the models, quantile residual diagnostics, estimation of generalized impulse response functions, and other numerical analysis are carried out with the CRAN distributed R package gmvarkit \citep{gmvarkitnormal} that accompanies this paper. The R package gmvarkit  also contains the dataset studied in the empirical application to facilitate reproduction of our results. The estimation is based on the exact log-likelihood function. For evaluating the adequacy of the models, we employ quantile residual diagnostics in the framework of \cite{Kalliovirta+Saikkonen:2010}  \citep[see also the related paper by][for discussion on quantile residual based model diagnostics in a univariate setting]{Kalliovirta:2012}. For a correctly specified GMVAR model, the empirical counterparts of the quantile residuals are asymptotically independent with multivariate standard normal distributions and can hence be used for graphical analysis in a similar manner to the conventional Pearson residuals \citep[Lemma 3]{Kalliovirta+Saikkonen:2010}.\footnote{\cite{Kalliovirta+Saikkonen:2010} also propose formal diagnostic tests for testing normality, autocorrelation, and conditional heteroskedasticity of the quantile residuals. The tests take into account the uncertainty about the true parameter value and can be calculated based on the observed data or by employing a simulation procedure for better size properties. We found these tests very forgiving without the simulation procedure and quite conservative without it. For instance, when taking into account the first four lags in the autocorrelation test, without the simulation procedure our GMVAR($3,2$) model obtains the $p$-value $0.999$, while with the simulation procedure, using a sample of length $10000$, the $p$-value is $0.000$. We therefore rather employ graphical diagnostics and compare the statistical properties of the quantile residuals to the ones of four-variate IID standard normal process.}

We started by estimating linear Gaussian VARs with the autoregressive orders $p=1,...,12$, i.e., GMVAR($p,1$) models. BIC was minimized by the order $p=1$, HQIC by the order $p=2$, and AIC by the order $p=3$, suggesting that the appropriate autoregressive order is likely relatively small. Hence, we then estimated the two-regime GMVAR($p,2$) models with $p=1,...,4$. BIC was minimized by the order $p=1$ and HQIC and AIC by the order $p=2$. Graphical quantile residual diagnostics revealed the order $p=1$ clearly inadequate to capture the autocorrelation structure of the series, while also the order $p=2$ was somewhat inadequate (not shown). We therefore considered the order $p=3$, which we found adequate to capture the autocorrelation structure of the series (see Section~\ref{sec:adequacy}). We also considered the order $p=4$, but since it increased AIC from $p=3$, which was already found adequate to explain the autocorrelation structure, and the order $p=3$ was found suitable for the linear VAR as well, we preferred the more parsimonious GMVAR($3,2$) model. The log-likelihoods and values of the information criteria are presented in Table~\ref{tab:ic} for the discussed models.

\begin{table}%[t!]
\centering
\begin{tabular}{c c c c c c}
Model        & Log-lik  & BIC     & HQIC    & AIC  \\ 
\hline\\[-1.5ex]
GMVAR($1,1$) & $-4.404$ & $9.430$ & $9.191$ & $9.030$ \\
GMVAR($2,1$) & $-4.245$ & $9.444$ & $9.077$ & $8.831$ \\
GMVAR($3,1$) & $-4.160$ & $9.606$ & $9.112$ & $8.780$ \\
GMVAR($1,2$) & $-3.558$ & $8.381$ & $7.894$ & $7.568$ \\
GMVAR($2,2$) & $-3.276$ & $8.481$ & $7.740$ & $7.242$ \\
GMVAR($3,2$) & $-3.183$ & $8.958$ & $7.961$ & $7.292$ \\
GMVAR($4,2$) & $-3.113$ & $9.482$ & $8.230$ & $7.390$ \\
\hline
\end{tabular}
\caption{The log-likelihoods and values of the information criteria divided by the number of observations for the discussed GMVAR($p,M$) models.}
\label{tab:ic}
\end{table}

Table~\ref{tab:ic} shows that the GMVAR($3,2$) model has significantly smaller BIC, HQIC, and AIC than all of the linear VARs. According to graphical quantile residual diagnostics, the GMVAR($3,2$) model also explains the statistical characteristics  of the data more adequately than the linear VARs (see Section~\ref{sec:adequacy}; graphical diagnostics of the linear VARs are not shown for brevity). Hence, we find our GMVAR($3,2$) model superior to the linear VARs.

It is possible that the superior fitness is due to the accommodation of time-varying covariance matrix or intercepts and cannot be attributed to the time-varying autoregression (AR) matrices. To test whether this is the case, we estimated two additional GMVAR($3,2$) models. In the first one, we constrained the AR matrices and intercept parameters to be identical in both regimes, thereby allowing for time-varying covariance matrix only. In the second one, we constrained the AR matrices to be identical in both regimes, thereby allowing for time-varying intercepts and covariance matrix only. Because the constrained models are nested to the GMVAR($3,2$) model and the maximum likelihood estimator has the conventional asymptotic distribution under the conventional assumptions \cite[Theorem 3]{Kalliovirta+Meitz+Saikkonen:2016}, we can test the validity of the constraints with a likelihood ratio test. The likelihood ratio test produces the $p$-value $0.011$ for the former type of constraint and the $p$-value $0.007$ for the latter type of constraints, thus, rejecting both constraints the $5\%$ level of significance and the latter type of constraints with the $1\%$ level of significance.\footnote{For robustness, we also estimated the constrained models with autoregressive orders $p=1,2,4$ (and $M=2$) and tested the validity of the constraints in these models with the likelihood ratio test. Both types of constraints were rejected for the $p=2,4$ models with $p$-values less than $0.0008$, but the GMVAR($1,2$) model accepted constraining the AR matrices to be identical in both regimes with the $p$-value $0.3$ and rejected constraining both AR matrices and intercepts with the $p$-value $0.005$. The rejection of the constraints does not, hence, seem particularly sensitive to the choice of $p$. Notably, the validity of the likelihood ratio test requires that the unrestricted model is correctly specified, so the tests that do not use the correct autoregressive order $p$ do not produce reliable result. If one of the orders $M=2$ and $p=2,3,4$ is correct, the constraints are, nevertheless, rejected.}

According to thee small $p$-values, it seems likely that also the AR matrices vary in time. But since the $p$-values were not particularly small, we studied the generalized impulse response functions of the constrained GMVAR($3,2$) models as well. We found that if the AR matrices and intercepts are constrained identical in both regimes (a linear VAR with two volatility regimes), small and large as well as contractionary and expansionary monetary policy shocks induce a long-run price puzzle in both regimes. If only the AR matrices are constrained to be identical in both regimes, small and large contractionary monetary policy shocks induce a medium- to long-run price puzzle and expansionary shocks medium-run price puzzle in the unstable inflation regime. But the GIRFs did not change much in the stable inflation regime (and thus display a long-run price puzzle for contractionary shocks in this regime as well). The GIRFs of the constrained models are not shown for brevity. That is, enforcing linearity to the autoregressive dynamics appears to make the price puzzle worse.\footnote{For the constrained models, the monetary policy shock was identified with the constraints described in Section~\ref{sec:identmone} similarly to the unconstrained model, as the unrestricted impact effects were similar in sign and magnitude to the unconstrained model. Also the estimated mixing weights were quite similar: one of the regimes prevailed in the 70's, 80's, the Financial crisis, and COVID-19 crisis, but also short periods during other times. For the ease of communication, we hence refer to them as stable inflation regime and unstable inflation regime similarly to the unconstrained model.} 

For comparison, we also considered a Cholesky identified Gaussian SVAR model with the autoregressive order $p=3$ (as suggested by AIC) and the interest rate variable ordered last. This model did not only display a  long-run price puzzle but it also displayed a short-run output puzzle, i.e., the response of GDP was positive (in the point estimate) before it became negative as a response to a contractionary monetary policy shock. A Cholesky SVAR with the order $p=4$ shows permanent decrease in the price level after roughly 9 years from the impact in response to a contractionary monetary policy shock. But the response of GDP still has the wrong sign  at the point estimate in the period after the impact (not shown). 

\subsection{Adequacy and characteristics of the selected model}\label{sec:adequacy}
\begin{figure}[H]
    \centerline{\includegraphics[width=\textwidth - 2cm]{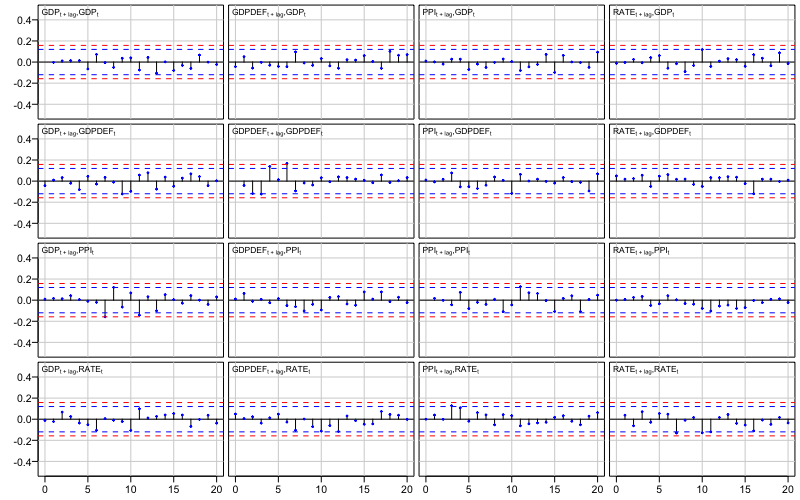}}
    \caption{Auto- and crosscorrelation functions of the quantile residuals of the fitted GMVAR($3,2$) model for the lags $0,1,...,20$.  The lag zero autocorrelation coefficients are omitted, as they are one by convention. The blue dashed lines are the $95\%$ bounds $\pm 1.96/\sqrt{T}$ ($T=267$ as the first $p=3$ observations were used as the initial values) for autocorrelations of IID observations, whereas the red dashed lines are the corresponding $99\%$ bounds $\pm 2.58/\sqrt{T}$. These bounds are presented to give an approximate perception on the magnitude of the correlation coefficients.}
\label{fig:acplot}
\end{figure}

In order to study the adequacy of our GMVAR($3,2$) model, we examine the quantile residual time series, sample auto- and crosscorrelation functions of the quantile residuals and squared quantile residuals,  and normal quantile-to-quantile plots. The sample auto- and crosscorrelation functions (presented in Figure~\ref{fig:acplot}) show that there is not much auto- or crosscorrelation in the quantile residuals.
The GDP deflator has some moderate sized autocorrelation coefficients (ACC) at small lags, but they are not very large. There are also moderate sized coefficients at larger lags in the crosscorrelation function of GDP and PPI as well as in the autocorrelation function of the interest rate variable. Nonetheless, given that in total of $316$ correlation coefficients are presented, some of them are expected to be moderate sized for an IID process as well.\footnote{Increasing the autoregressive order to $p=4$ reduces some of the ACCs of the GDP deflator's quantile residuals, but it does not help with the price puzzle.}

The sample auto- and crosscorrelation functions of the squared quantile residuals are presented in Figure~\ref{fig:acplot}. GDP, GDP deflator, and PPI each have at least one exceedingly large ACC in their autocorrelation functions, but ACCs of the interest rate variable are reasonable. There are also two exceedingly large coefficients at large lags in the crosscorrelation function of the interest rate variable and GDP deflator. Our GMVAR($3,2$) model is therefore clearly inadequate to capture the conditional heteroskedasticity in the series.

\begin{figure}[H]
    \centerline{\includegraphics[width=\textwidth - 2cm]{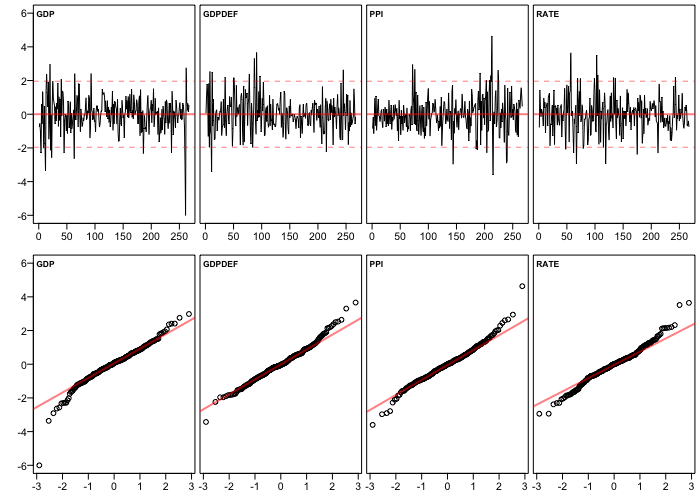}}
    \caption{Quantile residual time series and normal quantile-quantile-plots of the fitted GMVAR($3,2$) model.}
\label{fig:serqqplot}
\end{figure}

The quantile residual time series (the top panels of Figure~\ref{fig:serqqplot}) also show some heteroskedasticity and several outliers in the quantile residuals. There is a particularly large (marginal) quantile residual of the GDP in the beginning of the COVID-19 crisis, when the COVID-19 lockdown caused a fast and vast drop in the cycle. We do not view this large negative quantile residual of the GDP as an inadequacy, however, as the COVID-19 drop is known to be caused by an exceptionally large exogenous shock, and therefore large (quantile) residual is expected for a correctly specified model. The normal quantile-quantile-plots (the bottom panels of Figure~\ref{fig:serqqplot}) show that the marginal quantile residual distributions have excess kurtosis but are quite symmetric. The quantile residuals of GDP deflator seem slightly skewed to the right and GDP slightly to the left, however.

In our view, the overall adequacy of the model is decent enough for further analysis, particularly since autocorrelation structure of the data is captured reasonably well. Some of the conditional heteroskedasticity in the data remains unmodelled, which is not completely innocent because the mixing weights may depend on the volatility of the series. The unmodelled conditionally heteroskedasticy is not very extreme though: there is a single ACC of roughly the size $0.3$ in the autocorrelation functions of the squared quantile residuals of GDP, GDPDEF, and PPI, while almost all of the crosscrorrelation coefficients are of reasonable size. The excess kurtosis in the marginal distribution of the quantile residuals, in turn, does not seem particularly severe. Accommodating stronger forms of conditional heteroskedasticity and excess kurtosis by utilizing Student's $t$ distributed error terms similarly to \cite{Meitz+Preve+Saikkonen:2021} in the univariate setting is beyond the scope of this paper and left for future research \cite[see][]{Virolainen3:2021}.\footnote{The GMVAR model's capability to capture the conditional heteroskedasticity and marginal distribution of the series can be improved by adding a third regime. With $p=3$, however, the number of parameters in each regime is rather high: $62$ plus a mixing weight parameter for all but the last regime. In the most rare regime, this may be too much compared to the number of observations from that regime for any meaningful inference based on the estimates to take place. With smaller $p$, on the other hand, the model's capability to capture the autocorrelation structure of the series at larger lags might not be adequate. Because the three regime models are also tedious to estimate in practice, we focus on the two regime models.}

\begin{table}%[t!]
\centering
\begin{tabular}{c c c c c c c c c c}
               &  & \multicolumn{2}{c}{GDP} & \multicolumn{2}{c}{GDPDEF} & \multicolumn{2}{c}{PPI} & \multicolumn{2}{c}{RATE} \\
                & $\hat{\alpha}_m$ &$\hat{\mu}_{m,1}$ & $\hat{\sigma}_{m,1}^2$ & $\hat{\mu}_{m,2}$ & $\hat{\sigma}_{m,2}^2$  & $\hat{\mu}_{m,3}$ & $\hat{\sigma}_{m,3}^2$  & $\hat{\mu}_{m,4}$ & $\hat{\sigma}_{m,4}^2$ \\ 
\hline\\[-1.5ex]
Regime 1 & $0.56$ & $-0.20$ & $1.92$ & $0.71$ & $0.20$ & $0.72$ & $1.87$ & $4.44$ & $12.12$  \\
Regime 2 & $0.44$ & $-0.07$ & $7.83$ & $1.45$ & $0.95$ & $1.75$ & $10.96$ & $7.57$ & $34.28$  \\
\hline
\end{tabular}
\caption{Mixing weight parameter estimates ($\hat{\alpha}_m$) and marginal stationary means ($\hat{\mu}_{m,i}$) and variances ($\hat{\sigma}_{m,i}^2$) of the component series implied by the fitted GMVAR($3,2$) model for each of the regimes. }
\label{tab:regimes}
\end{table}

The estimated mixing weights of the two regimes are presented in the bottom panel of Figure~\ref{fig:seriesplot}. The second regime (red) mainly dominates during the periods of high inflation and interest rate in the 1970's and 1980's, after the collapse of Lehman Brothers in the Financial crisis until the end of 2009, and finally during the COVID-19 crisis from the third quarter of 2020 onwards. The first regime (blue) prevails when the second one does not: before 1970's, short periods during 1970's, and from the mid 1980's onwards but excluding the Financial crisis and the COVID-19 crisis (but including the first two quarters of 2020). Therefore, it appears that the second regime is mainly dominant when inflation has been high or volatile, while the first regime is dominant in more stable times.

The mixing weight parameters have the interpretation of being the unconditional probabilities for an observation being generated from each regime. For a correctly specified model, they should hence approximately reflect the proportions of observations generated from each regime. The first regime has a mixing weight parameter estimate $0.56$ (shown in Table~\ref{tab:regimes}), and it covers approximately $81\%$ of the series (approximated as the mean of the estimated mixing weights),  whereas the second regime has the implied mixing weight parameter estimate $0.44$ and it covers approximately $19\%$ of the series. The mixing weight parameter estimates are therefore somewhat disproportionate to the relative number of observations from each regime. This can be attributed to an estimation error (rather than misspecification), however, as the approximate standard error for the first regime's mixing weight parameter estimate is as high as $0.213$. Nonetheless, the mixing weight parameter estimates seem reasonable enough not to distort the generalized impulse response functions too much.  

Based on the model implied marginal stationary means and variances presented in Table \ref{tab:regimes}, neither of the regimes is particularly recessionary or expansionary, but the first regime has lower unconditional mean for the GDP, while the second one has much higher unconditional variance. Both regimes also prevail during recessions and expansions (see Figure~\ref{fig:seriesplot}). In the first regime, the GDP deflator has the (estimated) unconditional mean $0.71$, which implies long-run yearly inflation of approximately $2.9\%$, and unconditional variance $0.20$. In the second regime, the GDP deflator has the (estimated) unconditional mean $1.45$, which implies long-run yearly inflation of approximately $5.9\%$, and unconditional variance $0.95$. That is, the estimated long-run inflation is quite reasonable in the first regime, while it is excessive and volatile in the second regime. Also the commodity price inflation and the interest rate variable have much higher unconditional mean and variance the second regime than in the first regime. Based on the significantly higher unconditional means and variances of the inflation, commodity price inflation, and the interest rate variable, as well as on the timing of the dominance of the regimes (see Figure~\ref{fig:seriesplot} and the discussion above), we refer to the second regime as the unstable inflation regime. Accordingly, we refer to the first regime as the stable inflation regime. 

\begin{figure}[H]
    \centerline{\includegraphics[width=\textwidth - 2cm]{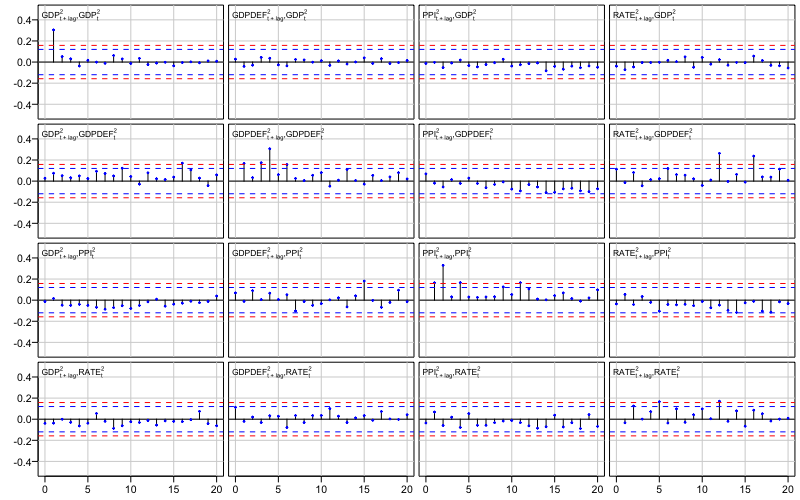}}
    \caption{Auto- and crosscorrelation functions of the squared quantile residuals of the fitted GMVAR($3,2$) model for the lags $0,1,...,20$.  The lag zero autocorrelation coefficients are omitted, as they are one by convention. The blue dashed lines are the $95\%$ bounds $\pm 1.96/\sqrt{T}$ ($T=267$ as the first $p=3$ observations were used as the initial values) for autocorrelations of IID observations, whereas the red dashed lines are the corresponding $99\%$ bounds $\pm 2.58/\sqrt{T}$. These bounds are presented to give an approximate perception on the magnitude of the correlation coefficients.}
\label{fig:chplot}
\end{figure}

\subsection{Estimates of the structural parameters}\label{sec:structparestimates}
The identification in %(\ref{eq:Wimposed}) 
Section~\ref{sec:identmone} produced the following estimates for the GMVAR($3,2$) model:
\small
\begin{equation}\label{eq:estimates2}
\hat{W}_2=\begin{bmatrix*}
\phantom{-}\boldsymbol{0.14} \ (0.054) & \phantom{-}\boldsymbol{0.22} \ (0.065) & \boldsymbol{0.45} \ (0.037) &           -0.12 \ (0.065) \\
          \boldsymbol{-0.20} \ (0.014) &           -0.05 \ (0.028) & \boldsymbol{0.07} \ (0.012) & \phantom{-}0\phantom{.00   (0.000)} \\
\phantom{-}0.00 \ (0.168) &           \boldsymbol{-1.03} \ (0.078) & \boldsymbol{0.47} \ (0.109) & -0.05 \ (0.068) \\
\phantom{-}0.03 \ (0.028) & \phantom{-}0.03 \ (0.041) & \boldsymbol{0.17} \ (0.044) & \phantom{-}\boldsymbol{0.30} \ (0.026) \\
\end{bmatrix*},
\ \
\hat{\lambda}_2 = \begin{bmatrix}
 \boldsymbol{\phantom{1}1.08} \ (0.227) \\
 \boldsymbol{\phantom{1}3.02} \ (0.640) \\
           \boldsymbol{11.05} \ (2.583) \\
           \boldsymbol{18.20} \ (4.398) \\
\end{bmatrix},
\end{equation}
\normalsize
where the ordering of the variables is $y_t=(GDP_t, GDPDEF_t, PPI_t, RATE_t)$ and approximate standard errors are given in parentheses next to the estimates. The estimates that deviate from zero by more than two times their approximate standard error are bolded.

\subsection{Individual GIRFs in the stable inflation regime}\label{sec:indgirfres}
\begin{figure}
    \centerline{\includegraphics[width=\textwidth - 2cm]{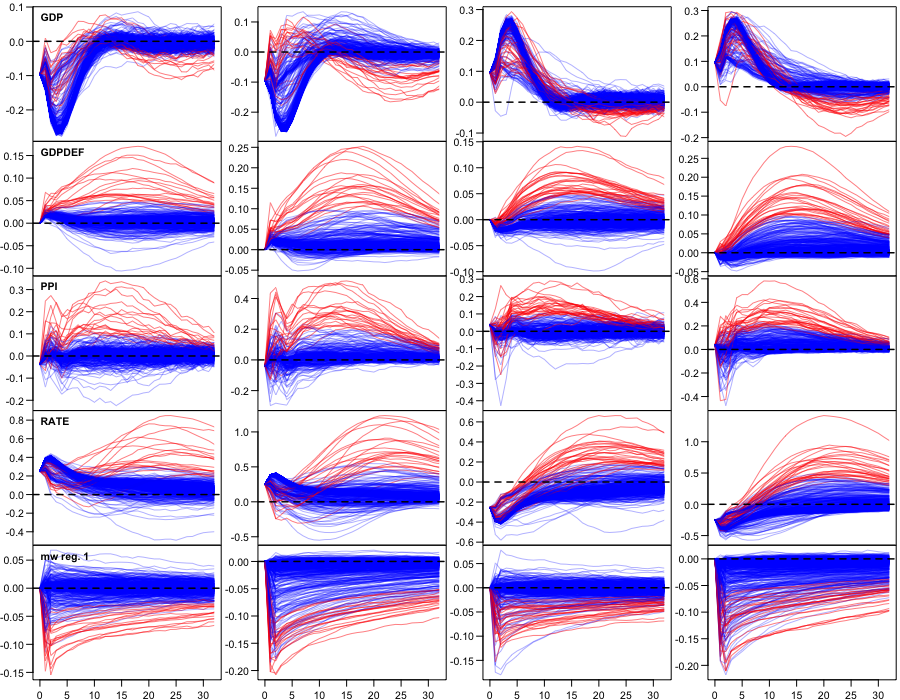}}
    \caption{Generalized impulse response functions of the identified monetary policy shock of the fitted GMVAR($3,2$) model. Each column presents $500$ GIRFs, each based on a random starting value drawn from the stationary distribution of the first regime and $2500$ Monte Carlo repetitions. The first (third) column present GIRFs to one-standard-error contractionary (expansionary) shocks and the second (fourth) column to two-standard-error contractionary (expansionary) shocks. From the top, the first panels show the response of the cyclical component of GDP, the second panels show the response of log-differenced implicit GDP deflator, the third panels show the response of log-differenced producer price index (all commodities), the fourth panels show the response of the interest rate variable, and the bottom panels show the response of the first regime's mixing weights. The GIRFs are scaled to correspond to $25$ basis point instantaneous increase (decrease) of the interest rate variable. The GIRFs that display (scaled) peak inflation greater than $5$ basis points for one-standard-error shocks and $10$ basis for two-standard-error shocks are colored red and the rest blue.}
\label{fig:indgirf}
\end{figure}

The confidence bounds of the GIRFs are relatively wide in the stable inflation regime (the first and second columns of Figure~\ref{fig:girfplot}), and they display some unexpected results such as prices rising as a response to a contractionary monetary policy shock, the interest variable overshooting significantly as a response to an expansionary monetary policy shock, and counterproductive response of the GDP after the initial expansion (or contraction). In order to investigate how these results appear in the model dynamics and to what extend they might be economically sensible, we have depicted $500$ individual GIRFs (each estimated based on $2500$ Monte Carlo repetitions) in each column of Figure~\ref{fig:indgirf} with the starting values generated from the stable inflation regime. 

The first (third) column of Figure~\ref{fig:indgirf} presents the GIRFs to a one-standard-error contractionary (expansionary) monetary policy shock and the second (fourth) column presents the GIRFs to a two-standard-error contractionary (expansionary) monetary policy shock. After estimating the GIRFs, they were scaled to correspond to a $25$ basis point increase (decrease) of the interest rate variable. The GIRFs that display (scaled) peak inflation greater than $5$ basis points for one-standard-error shocks and $10$ basis points for two-standard-error shocks are colored red and the rest blue. 

One of the unexpected observations in Section~\ref{sec:girfresults} is that (in the stable inflation regime) the prices seem to often rise in response to both contractionary and expansionary monetary policy shocks, particularly if the shock is large. From the perspective of the model dynamics, the reason is that from many of the starting values the monetary policy shock drives the economy towards the unstable inflation regime, as we next explain. The bottom row of Figure~\ref{fig:indgirf} shows that in the high inflation red GIRFs, the probability of the unstable inflation regime increases sharply in the period after the impact (while at impact the mixing weights are predetermined). This implies that the impact responses of the observable variables induce a greater probability of the unstable inflation regime, which then moves the observable variables in the following periods accordingly. Thus, the monetary policy shock drives the economy towards the unstable inflation regime (that has high long-run inflation), which in part causes an increase in inflation (and not the vice versa).

The red GIRFs in each column of Figure~\ref{fig:indgirf} show that the GIRFs exhibiting particularly high increase in inflation (and commodity price inflation) also display a persistent increase in the interest rate variable. Given the movements of the prices, the response of the interest rate variable is economically sensible when the Fed's endogenous response to high inflation is tight monetary policy. When the interest rate rises, inflation starts to finally decrease after several years from impact but so does the GDP. 

If the shock is contractionary (the first and second columns of Figure~\ref{fig:indgirf}), the GDP temporarily recovers in the red GIRFs relatively fast before persistently decreasing along with the rising interest rate. The temporary recovery of the GDP might be related to the higher unconditional mean of the GDP in the unstable inflation regime. Since there are only approximately $49$ observations from the unstable inflation regime (estimated as the sum of the mixing weights), the temporary recovery can possibly be attributed to estimation error. Also, particularly when the contractionary shock is large, from some of the initial values the GDP's response overshoots significantly to the positive side without the delayed contraction that the red GIRFs display. Further investigation revealed that these GIRFs are mostly the ones that display positive peak deflation and a response of the interest rate variable that overshoots to the negative side, and therefore, the expansion has an economic explanation through the expansionary monetary policy (not shown).

If the shock is expansionary (the third and fourth columns of Figure~\ref{fig:indgirf}), both high inflation red GIRFs and low inflation blue GIRFs display roughly same length expansions of the GDP. The high inflation red GIRFs in which the interest rate significantly overshoots to the positive side, however, display a delayed contraction of the GDP after the initial expansion. That is, particularly a large expansionary (but also a contractionary) shock drives the economy towards the unstable inflation regime, in part causing the high inflation. This results in significant monetary policy tightening, which is accompanied with a persistent contraction of the GDP. 

\end{appendices}

\end{document}